\documentclass{MySigAltFull}
\usepackage{textcomp}
\usepackage{graphics}
\usepackage{url}
\newcommand{\msg}[1]{{\small \textsf{#1}}}
\allowdisplaybreaks		

\begin{document}

\hyphenation{foot-hold foot-holds Ocean-store man-u-script man-u-scripts
high-er time-scale}

\title{Preserving Peer Replicas By Rate-Limited Sampled
Voting in LOCKSS\titlenote{This is an extended version of a shorter conference
paper~\cite{Maniatis2003lockssSOSP}.}}

\numberofauthors{3}
\author{
\alignauthor Petros Maniatis\titlenote{Now with Intel Research,
  Berkeley, CA.}\\
             \url{maniatis@cs.stanford.edu}\\
	     \vspace{.1in}
             David S. H. Rosenthal\\
	     \url{dshr@stanford.edu}
\alignauthor Mema Roussopoulos\titlenote{Now with Harvard University,
             Cambridge, MA.}\\
             \url{mema@cs.stanford.edu}\\
	     Mary Baker\titlenote{Now with HP Labs,
             Palo Alto, CA.}\\
	     \url{mgbaker@cs.stanford.edu}\\
	     \vspace{.2in}
	     \affaddr{Stanford University}\\
	     \affaddr{Stanford, CA 94305}
\alignauthor TJ Giuli\\
             \url{giuli@cs.stanford.edu}\\
	     \vspace{.1in}
	     Yanto Muliadi\\
	     \url{cylee@stanford.edu}
}

\maketitle

\begin{abstract}
The LOCKSS project has developed and deployed in a world-wide test a
peer-to-peer system for preserving access to journals and other archival
information published on the Web.  It consists of a large number of
independent, low-cost, persistent web caches that cooperate to detect
and repair damage to their content by voting in ``opinion polls.''
Based on this experience, we present a design for and simulations of a
novel protocol for voting in systems of this kind.  It incorporates rate
limitation and intrusion detection to ensure that even some
very powerful adversaries attacking over many years have only a small
probability of causing irrecoverable damage before being detected.

{
\tiny
\begin{verbatim}
$Id: SOSP2003-long.tex,v 1.19 2003/10/17 03:40:01 maniatis Exp $
\end{verbatim}
}

\end{abstract}


\section{Introduction}
\label{sec:introduction}

Academic publishing is migrating to the Web~\cite{Mogge1999},
forcing the libraries that pay for
journals to transition from purchasing copies of the
material to renting access to the publisher's copy~\cite{Keller2003}.
Unfortunately,
rental provides no guarantee of long-term access.
Librarians consider it one of their responsibilities to provide
future readers access to important materials.
With millennia of experience with physical documents, they have
techniques for doing so:
acquire lots of copies of the document,
distribute them around the world,
and lend or copy them when necessary to provide access.

In the LOCKSS\footnote{LOCKSS is a trademark of Stanford University.}
program (Lots Of Copies Keep Stuff Safe), we model the physical document
system and apply it to Web-published academic journals,
providing tools for libraries to take custody of the
material to which they subscribe, and to cooperate with
other libraries to preserve it and provide access.  The LOCKSS approach
deploys a large number of independent, low-cost, persistent web caches
that cooperate to detect and repair damage by voting in ``opinion
polls'' on their cached documents.  The initial version of the
system~\cite{Rosenthal2000} has been under test since 1999
at about 50 libraries world-wide, and is expected to be in
production use at many more libraries in 2004.  Unfortunately, the
protocol now in use does not scale adequately, and analysis of the
first design for a revised protocol~\cite{Michalakis2003} showed it
to be insufficiently resistant to attack.

In this work, we present a design for and simulations of a new
peer-to-peer opinion poll protocol that addresses these scaling
and attack resistance issues. We plan to migrate it to
the deployed system shortly.
The new protocol is based on our experience with the deployed LOCKSS
system and the special characteristics of such a long-term large-scale application.
Distributed digital preservation, with its time horizon of many decades
and lack of central control, presents both unusual requirements, such as
the need to avoid long-term secrets like encryption keys, and unusual
opportunities, such as the option to make some system operations
inherently very time-consuming without sacrificing usability.

Digital preservation systems must resist both random failures and
deliberate attack for a long time.  Their ultimate success can be
judged only in the distant future.  Techniques for evaluating their
design must necessarily be approximate and probabilistic; they share
this problem with encryption systems.  We attempt to evaluate our design
in the same way that encryption systems are evaluated, by estimating the
computational effort an adversary would need to achieve a given
probability of the desired result.  In an encryption system, the desired
result is to recover the plaintext. In our case, it is to have the
system deliver a corrupt copy of a document.  These estimates can be
converted to monetary costs using technology cost curves, and thus
compared to the value of the plaintext or document at risk.

We introduce our design principles and the deployed test system, then
describe our new protocol and the reasons for the design decisions we
made.  We analyze some attacks its adversary can mount, then describe
simulations of the system and three types of such attack, aimed at
undetected direct corruption of a document, at discrediting the
system by causing alarms, and at slowing the system down.

Our simulations show a system that resists for decades an adversary
capable of unlimited sustained effort, by preventing him from applying
effectively more effort to the system than his victims do.  Even assuming
that an implementation flaw hands an adversary instantaneous control of
one-third of the peers, his sustained effort can increase the
probability of a reader seeing a damaged copy by no more than a further
3.5\%, in the worst case.  The system has a high probability of detecting an attack that can cause
permanent damage before that damage becomes irreversible, while
producing very few false positives due to random faults.

We believe this protocol and its underlying
principles are novel and will prove useful in the design of other
long-term large-scale applications operating in hostile environments.

\section{Design Principles}
\label{sec:designPrinciples}

Digital preservation systems have some unusual features.  First, such
systems must be very cheap to build and maintain, which precludes
high-performance hardware such as RAID~\cite{Patterson1988}, or complicated administration.
Second, they need not operate quickly.  Their purpose is to prevent
rather than expedite change to data.  Third, they must function properly
for decades, without central control and despite possible interference
from attackers or catastrophic failures of storage media such as fire or
theft.  These features, combined with our experience building and
maintaining other large-scale distributed systems, lead to the very
conservative design principles we use:

\emph{Cheap storage is unreliable.}  We assume that in our timescale no
cheap and easy to maintain storage is
reliable~\cite{ConservationOnline}.  Note that write-once media are at
least as unreliable as disks, eliminating alternate designs dependent on
storing documents or their hashes on CD-R (in our current deployment the
CD-R containing the peer software is the cause of the vast majority of
errors).

\emph{No long-term secrets.}  Or, to quote Diffie~\cite{diffieQuote},
``The secret to strong security: less reliance on secrets.''  Long-term
secrets, such as private keys, are too vulnerable for our application.
These secrets require storage that is effectively impossible to
replicate, audit, repair or regenerate.  Over time they are likely to leak;
recovering from such leakage is extraordinarily
difficult~\cite{Davis1996,Venema1996}.

\emph{Use inertia.}  The goal of a digital preservation system is to
prevent change.  Some change is inevitable, and the system must repair
it, but there is never a need for rapid change.  A system that fails
abruptly, without warning its operators in time for them to take
corrective action and prevent total failure~\cite{Staniford2002}, is not
suitable for long-term preservation.  Rate limiting has proved useful in
other areas~\cite{Williamson2002}; we can exploit similar techniques
because we have no need for speed.

\emph{Avoid third-party reputation.}  Third-party reputation information
is subject to a variety of problems, especially in the absence of a
strong peer identity.  It is vulnerable to slander and subversion of
previously reliable peers.  If evidence of past good behavior is
accumulated, an attacker can ``cash in'' a history of good behavior in
low-value interactions by defecting in a single high-value
interaction~\cite{EBayScam}.

\emph{Reduce predictability.}
Attackers predict the behavior of their victim to choose tactics.
Making peer behavior depend on random combinations of external inputs
and internal state reduces the accuracy of these predictions.

\emph{Intrusion detection is intrinsic.}
Conventional intrusion detection systems are extrinsic to the application
being protected.
They have to operate with less than full information about it,
and may themselves become a target.
Systems with bimodal behavior~\cite{Birman1999}
can provide intrinsic intrusion detection by surrounding good
states with a ``moat'' of forbidden states
that are almost never reachable in the absence of an attack, and that
generate an alarm.

We believe this mechanism to be fundamentally more robust than layering
an intrusion detection system on top of an application;
it does however share with conventional intrusion detection systems
the notion that repelling
attacks on the system must be a cooperative effort between the
software and the humans responsible for it.

\emph{Assume a strong adversary.}
The LOCKSS system preserves e-journals that have intrinsic value
and contain information that powerful interests might want
changed or suppressed.
Today,
a credible adversary is an Internet worm whose payload attacks the
system using tens of thousands of hosts.
We must plan for future, more powerful attacks.

The LOCKSS design is very conservative, appropriately so for a
preservation system.  Our goal is to apply these principles to the
design to achieve a high
probability that even a powerful adversary fails to cause
irrecoverable damage without detection.

\section{The Existing LOCKSS System}
\label{sec:background}

The LOCKSS system models librarians' techniques for physical documents
to preserve access to e-journals, by making it appear to a library's
patrons that pages remain available at their original URLs even if
they are not available there from the publisher.  We thus preserve
access to the material via common techniques such as links, bookmarks,
and search engines.  To do this, participating libraries run persistent
web caches that:
\begin{itemize}

\item \emph{collect} by crawling the journal web-sites to pre-load
themselves with newly published material,

\item \emph{distribute} by acting as a limited proxy cache for the library's
local readers,
supplying the \emph{publisher's} copy if it is available and the local copy
otherwise,

\item \emph{preserve} by cooperating with other caches that hold the same
material to detect and repair damage.

\end{itemize}
Caches cooperate by participating in ``opinion polls'' in a
peer-to-peer network.
In each,
a sample of peers votes on the hash of a specified part of
the content.

Polls provide peers with confidence in content \emph{authenticity} and
\emph{integrity}.  Journal publishers do not currently sign the material
they distribute, they do not provide a manifest describing the 
files forming a given paper, issue or volume, and the crawling process
is unreliable.  Furthermore, no completely reliable long-term storage
medium is available.  Catastrophic failures such as fire, theft, and
hacker break-in can wipe out or alter any storage medium without possibility of
recovery.  Evidence that many peers independently obtained and agree
with each other on the material is the best available guarantee that
content is authentic and correctly preserved.

Peers vote on large \emph{archival units} (AUs),
normally a year's run of a journal.
Because each peer holds a different set of AUs,
the protocol treats each AU independently.
If a peer loses a poll on an AU,
it calls a sequence of increasingly specific partial
polls within the AU to locate the damage.
Other peers cooperate with the damaged peer if they remember it
agreeing with them in the past about the AU, by offering it a good
copy, in the same way they would for local readers.

This mechanism defends against two important problems endemic to
peer-to-peer systems: \emph{free-loading} and \emph{theft}.  First, the
only benefit a peer obtains from the system is a repair, and to obtain
it the peer must have participated in the past, which precludes free-loading.
Second, a peer only supplies material to a peer that proved in the past
that it had that material, so the system does not \emph{increase} the
risk of theft.  In this way LOCKSS peers provide a distributed, highly
replicated, self-healing store of data that does not materially increase
the risks that publishers already run.  This is important; under the
DMCA~\cite{EFF-DMCA} publishers must give permission for libraries to
preserve their material.

Library budgets are perennially inadequate~\cite{ARLstats}.  To be
effective, any digital preservation system must be affordable in the
long term.  Minimizing the cost of participating in the LOCKSS system is
essential to its success, so individual peers are built from low-cost,
unreliable technology.  A generic PC with three 180GB disks currently
costs under \$1000 and would preserve about 210 years of the largest
journal we have found (the \emph{Journal of Biological Chemistry}) for a
worst-case hardware cost of less than \$5 per journal/year.  This is
equivalent to less than 1{\textcent} per SOSP proceedings.

Using these generic PCs we can build a system with acceptable
performance.  If peers check each AU every three months and split their
time equally between calling polls and voting in polls called by others,
each peer has 45 days in which to call one poll for each of its AUs.  If
there are 210 AUs, each poll should last about 5 hours.  With our new
protocol, this size of AU costs the caller about 1040 seconds for each
peer it invites to vote (Section~\ref{sec:simulationEnvironment}).  Each
poll could thus involve about 17 peers, more than in the current tests.

Peers require little administration~\cite{Rosenthal2003b}, relying on
cooperation with other caches to detect and repair failures.  There is
no need for off-line backups on removable media.  Creating these
backups, and using them when readers request access to data, would
involve excessive staff costs and latencies beyond a reader's attention
span~\cite{UsabilityGuidelines}.

\section{The New Opinion Poll Protocol}
\label{sec:protocol}

In this section we outline, describe and justify our new LOCKSS opinion
poll protocol.  We give an overview of the protocol and introduce
relevant apparatus and notation before specifying the protocol in more
detail in Section~\ref{sec:protocol:specification}.  In
Section~\ref{sec:protocol:analysis}, we distill the main techniques we
employ in our design and explain how and why the protocol applies them.

To simplify the analysis of the new protocol, we abstract the relevant
features of the existing system.  We consider a population of peers
preserving a copy of a single AU, obtained from a publisher who is no
longer available.  We ignore the divide-and-conquer search for damage
in a real, multi-file journal.  Each peer uses one of a number of
independent implementations of the LOCKSS protocol to limit
common-mode failures.  Each peer's AU is subject to the same low rate
of undetected random damage.

While a \emph{peer} is any node that participates with
benign or malicious intent in the LOCKSS protocol, we make the
following distinctions between different types of peers in the rest
of this paper:
\begin{itemize}
\item A \emph{malign} peer is part of a conspiracy of peers attempting
to subvert the system.
\item A \emph{loyal} peer is a non-malign peer, i.e., one that follows
the LOCKSS protocol at all times.
\item A \emph{damaged} peer is a loyal peer with a damaged AU.
\item A \emph{healthy} peer is a loyal peer with the correct AU.
\end{itemize}

The overall goal of our design is that there be a high probability
that loyal peers are in the healthy state despite failures and attacks, and
a low probability that even a powerful adversary can damage a
significant proportion of the loyal peers without detection.

A LOCKSS peer calls opinion polls on the contents of an AU it holds at
a rate much greater than any anticipated rate of random
damage.  It invites into its poll a small subset of the
peers it has recently encountered, hoping they will offer votes on
their version of the AU.  Unless an invited peer is busy, it computes
a fresh digest of its own version of the AU, which it returns in a
vote.  If the caller of the poll receives votes that overwhelmingly
agree with its own version of the AU (a \emph{landslide win}), it is
satisfied and waits until it has to call a poll again.  If it receives
votes that overwhelmingly disagree with its own version of the AU (a
\emph{landslide loss}), it repairs its AU by fetching the copy of
a voter who disagreed, and re-evaluates the votes, hoping now to
obtain a landslide win for its repaired AU.  If the result of the poll
justifies neither a landslide win nor a landslide loss (an
\emph{inconclusive} poll), then the
caller raises an alarm to attract human attention
to the situation.

The protocol supports two roles for participating peers (see
Section~\ref{sec:protocol:specification}).  First, the \emph{poll
initiator} calls polls on its own AUs and is the sole beneficiary of
the poll result.  Second, the \emph{poll participant} or \emph{voter}
is a peer who is invited into the poll by the poll initiator and who
votes if it has the necessary resources.  A voter need
not find out the result of a poll in which it votes.  Poll
participants for a given poll are divided into two groups: the
\emph{inner circle} and the \emph{outer circle}.  Inner
circle participants are chosen by the poll initiator from those peers
it has already discovered.  The initiator decides the outcome of the poll
solely on inner circle votes.  Outer circle participants are chosen by
the poll initiator from peers nominated by inner circle voters.
The initiator uses outer circle votes to perform discovery, i.e., to
locate peers that it can invite into future inner circles for its
polls.

\begin{figure}
\centerline{\includegraphics{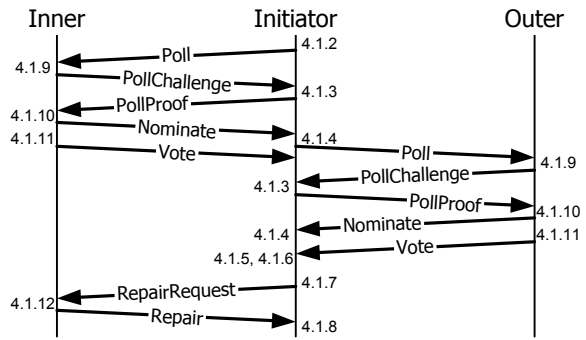}}
\caption{The protocol messages exchanged during a poll between the
poll initiator and poll participants.  Time flows from top to bottom.
Next to each phase in the execution of the protocol, we give the
section number that provides the pertinent description.}
\label{fig:Messages}
\end{figure}

LOCKSS peers communicate in two types of exchanges (see
Section~\ref{sec:protocol:specification}).  First, a poll
initiator uses unicast datagrams to communicate with the peers
it invites to arrange participation and voting in the poll.
Second, a poll initiator may contact its voters to request a
repair for its AU using a bulk transfer protocol.  In both
cases, communication is encrypted via symmetric session keys, derived
using  Diffie-Hellman key exchanges~\cite{Diffie1976} between the poll
initiator and each of its participants.  After the poll, session keys
are discarded.
Figure~\ref{fig:Messages} shows a
typical message exchange between a poll initiator and its inner and
outer circles.

The LOCKSS opinion poll protocol requires both
poll initiators and voters to expend provable
computational effort~\cite{Dwork1992} in amounts
related to underlying system operations (hashing of an AU),
as a means of limiting Sybil attacks~\cite{Douceur2002}.
We describe in Section~\ref{sec:protocol:analysis:effortSizing}
how these amounts are determined,
and how proofs of effort are constructed, verified, and used.
In the protocol description below we simply refer to the generation
and verification of effort proofs.

In the remainder of this paper we use the following notation for system
parameters:
\begin{description}
\item[$A$] Maximum number of discredited challenges allowed in a poll
(Section~\ref{sec:protocol:specification:initiation})
\item[$C$] Proportion of the reference list refreshed using friends at
every poll (\emph{churn factor} in
Section~\ref{sec:protocol:specification:referenceListUpdate})
\item[$D$] The maximum number of votes allowed to be in the minority of
  a poll (Section~\ref{sec:protocol:specification:tabulation})
\item[$E$] Maximum age of unused reference list entries
(Section~\ref{sec:protocol:specification:referenceListUpdate})
\item[$I$] Number of outer circle nominations per inner circle
participant
(Section~\ref{sec:protocol:specification:effortVerification})
\item[$N$] Number of inner-circle peers invited into a poll
(Section~\ref{sec:protocol:specification:initiation})

\item[$Q$] Number of valid inner-circle votes required to conclude
a poll successfully (\emph{quorum})
(Section~\ref{sec:protocol:specification:initiation})

\item[$R$] Mean interval between two successive polls called by a peer on the same
AU (Section~\ref{sec:protocol:specification:bootstrapping})
\end{description}
and for convenience variables:
\begin{description}
\item[$L$] Number of loyal voters in the inner circle
(Section~\ref{sec:adversary:strategies:stealth})

\item[$M$] Number of malign voters in the inner circle
(Section~\ref{sec:adversary:strategies:stealth})

\item[$V$] Number of inner-circle peers whose vote is received and
verified to be valid
(Section~\ref{sec:protocol:specification:tabulation})

\end{description}

\subsection{Detailed Description}
\label{sec:protocol:specification}

In this section, we present in detail how the opinion poll protocol
works.  In Section~\ref{sec:protocol:analysis}, we explain the reasoning
behind our major design decisions.

Each peer maintains two peer lists for every AU it holds: the
\emph{reference list}, which contains information about other LOCKSS
peers it has recently encountered; and the \emph{friends list}, which
contains information about LOCKSS peers with whose operators or
organizations the peer has an out-of-band relationship.  A peer
maintains for every AU it holds a \emph{poll counter} that records the
number of polls the peer has called on that AU since first acquiring it.

Reference list entries have the form [\emph{peer IP address},
\emph{time inserted}]. They are added to or removed from the list by
the protocol.  The value of the \emph{time inserted} field is set to
the value of the poll counter at the time the entry is inserted into
the reference list.  Friends list entries contain only a peer
IP address.  They are added to or removed from the list by the peer's
operator, as his affiliations with other institutions change over
time.

A peer who is not in the process of calling a poll for an AU also
maintains a \emph{refresh timer} for that AU.  When the timer expires,
the peer calls a new poll for the AU
(Section~\ref{sec:protocol:specification:initiation}).

In what follows, we describe in detail the different protocol steps,
including bootstrapping
(Section~\ref{sec:protocol:specification:bootstrapping}), the poll
initiator's point of view
(Sections~\ref{sec:protocol:specification:initiation} to
\ref{sec:protocol:specification:referenceListUpdate})
and the poll participant's point of view 
(Sections~\ref{sec:protocol:specification:solicitation} to
\ref{sec:protocol:specification:repairSolicitation}).

\subsubsection{Bootstrapping}
\label{sec:protocol:specification:bootstrapping}

When a peer first enters a LOCKSS network for a given AU, or when
it reinitializes after a failure, it copies all
entries from its current friends list into its reference list, and sets
its refresh timer with a random expiration time with mean value $R$.  In
our simulations, we choose this random value uniformly from an
interval centered at $R$.

\subsubsection{Poll Initiation}
\label{sec:protocol:specification:initiation}

To call a new poll on an AU, a LOCKSS peer chooses a fresh, random poll
identifier and $N$ random peers from its reference list, which it
inserts into its \emph{inner circle list}.  For each inner circle peer,
the poll initiator chooses a fresh, random Diffie-Hellman public key,
and sends that peer a \msg{Poll} message, of the form [\emph{Poll ID},
\emph{DH Public Key}].  Then the initiator waits for \msg{PollChallenge}
messages from the invited inner circle peers (see
Section~\ref{sec:protocol:specification:solicitation}) and sets a
\emph{challenge timer} to stop waiting.

The initiator removes from its inner circle list those peers who respond
with a negative \msg{PollChallenge} message, those who do not respond by
the time the challenge timer expires, and those from whom the initiator
receives multiple \msg{PollChallenge} messages with conflicting contents.

Peers removed because of conflicting \msg{PollChallenge} messages are
said to be \emph{discredited}.  A discredited peer may be
self-inconsistent because of a local fault; alternatively, it may be the
victim of a spoofer located near it or near the poll initiator.  Either
way, the initiator cannot tell the intended from the faulty or malicious
\msg{PollChallenge} messages, so it removes discredited peers from the
poll.  If the initiator discredits more than $A$ peers 
in a poll, it suspects a local spoofer and raises a spoofer alarm (see
Section~\ref{sec:protocol:specification:alarms}).

For all inner circle peers who send a valid, affirmative challenge, the
initiator computes the provable effort for its poll
invitation (Section~\ref{sec:protocol:specification:effort}).  If the
initiator ends up with fewer inner circle peers than $Q$, the minimum
required, it invites additional peers into the poll via more \msg{Poll}
messages, or aborts the poll if it has no more peers in its reference
list.

\subsubsection{Poll Effort}
\label{sec:protocol:specification:effort}

For every received affirmative \msg{PollChallenge} message, the initiator
produces some computational effort that is provable via a \emph{poll
effort proof} (see Section~\ref{sec:protocol:analysis:effortSizing}).
The effort and its proof are cryptographically derived from the poll
identifier and the potential voter's challenge.  The
initiator returns this poll effort proof to the sender of the associated
\msg{PollChallenge} message within a \msg{PollProof} message of the form
[\emph{Poll ID}, \emph{poll effort proof}], encrypted using the session
key.

The initiator also sends \msg{PollProof} messages to poll participants
who responded to the initial invitation with a negative
\msg{PollChallenge}.  The initiator need not expend computational effort
for negative challenges; it can use a random value as 
the poll effort proof (see Section~\ref{sec:protocol:analysis:obfuscation}).

After sending all \msg{PollProof} messages, the initiator
waits for \msg{Nominate} messages
(Section~\ref{sec:protocol:specification:effortVerification}) and sets
the \emph{nomination timer} to stop waiting.  When all \msg{Nominate}
messages arrive or the timer expires, the initiator forms its outer
circle.

\subsubsection{Outer Circle Invitation}
\label{sec:protocol:specification:outerCircle}

The initiator discovers new peers that maintain the same AU by forming
an outer circle based on the \msg{Nominate} messages returned from its
inner circle poll participants
(Section~\ref{sec:protocol:specification:effortVerification}).
Discovery is important when the reference list is short (close to
$N$), but less necessary when the reference list is long.  Therefore,
the initiator picks an outer circle size that, when added to its
current reference list, would achieve a target reference list size (in
our simulations $3 \times N$).

To form its outer circle, the initiator removes from every nomination
list peers already contained in its reference list, and then it chooses
an \emph{equal} number of peers from every nomination list at random for
its \emph{outer circle list}; as a result, every inner circle nominator
affects the outer circle equally.  The initiator invites outer circle
peers into the poll in a manner identical to inviting the inner circle,
as outlined in Sections~\ref{sec:protocol:specification:initiation} and
\ref{sec:protocol:specification:effort}.  \msg{Nominate} messages from
outer circle participants are ignored.

The initiator starts collecting \msg{Vote} messages once
it has sent its inner circle \msg{PollProof} messages
(Section~\ref{sec:protocol:specification:effort}).
As soon as it finishes the construction of poll effort proofs for
the outer circle, it sets a \emph{vote timer} to stop collecting
\msg{Vote} messages.
When all expected \msg{Vote} messages have arrived, or the vote timer has
expired, the initiator verifies the votes.

\subsubsection{Vote Verification}
\label{sec:protocol:specification:voteVerification}

Vote verification deems votes to be one of \emph{invalid}, \emph{valid
but disagreeing} with the initiator's AU, or \emph{valid but agreeing}
with the initiator's AU.  Votes are constructed in rounds
(Section~\ref{sec:protocol:specification:voteConstruction}) and
are thus verified in rounds.

In each such round, the initiator
verifies the proof of computational effort included in the \msg{Vote}
message for the corresponding voting round.
If the proof is incorrect, the initiator deems the vote invalid and
verification stops.  Otherwise, if the vote has yet to be deemed
disagreeing, the initiator hashes the proof with the corresponding portion
of its own copy of the AU;
if the result
does not match the hash in the vote, the vote is declared disagreeing.
The initiator skips hashing the AU if it has already deemed the vote disagreeing
and uses the values in the \msg{Vote} message to proceed with validity verification, instead.

If all proofs of effort are correct, the initiator deems the vote valid.
If all AU hashes match, the initiator deems the vote agreeing, and
disagreeing otherwise.  Invalid votes result in the removal of the
offending voter from the poll (inner or outer circle), and from the
initiator's reference list, since they indicate fault or malice.
When the initiator has verified all received \msg{Vote} messages,
it tabulates the results.

\subsubsection{Vote Tabulation}
\label{sec:protocol:specification:tabulation}

The poll initiator tabulates the valid votes from the inner circle
to determine whether its AU replica is correct.
If the number $V$ of valid inner circle votes is greater than or equal
to the quorum $Q$, then the participant acts as follows:
\begin{itemize}
\item \emph{Agreeing votes are no more than $D$}.  The poll is a
landslide loss. The initiator considers its current AU copy damaged
and repairs it (Section~\ref{sec:protocol:specification:repair}).
\item \emph{Agreeing votes are at least $V-D$}.  The initiator considers
its current copy the prevailing one (landslide win).  This is the only
way in which an opinion poll \emph{concludes successfully}.  The
initiator updates its reference list
(Section~\ref{sec:protocol:specification:referenceListUpdate}) and
schedules another poll at a random future time uniformly centered at
$R$.
\item \emph{Agreeing votes are more than $D$ but fewer than $V-D$}.  The
initiator considers the poll inconclusive and raises an alarm
(Section~\ref{sec:protocol:specification:alarms}).
\end{itemize}
If the initiator has been unable to accumulate $Q$
valid votes from its inner circle, then it does not make a decision on
its AU; it updates its reference list
(Section~\ref{sec:protocol:specification:referenceListUpdate}), and
immediately calls another poll.  If it has failed to obtain $Q$ votes in a
poll on this AU for a long time, the initiator raises an inter-poll
interval alarm (Section~\ref{sec:protocol:specification:alarms}).

\subsubsection{Repair}
\label{sec:protocol:specification:repair}

If the initiator decides that its AU is damaged, it picks at random one
of the disagreeing inner circle voters and sends it an encrypted
\msg{RepairRequest} message containing the poll identifier.  If it
receives a \msg{Repair} message (see
Section~\ref{sec:protocol:specification:repairSolicitation}), the
initiator re-verifies any disagreeing votes given the new AU
(Section~\ref{sec:protocol:specification:voteVerification}) and
re-tabulates the results
(Section~\ref{sec:protocol:specification:tabulation}).  If it does not
receive a \msg{Repair} message, it picks another disagreeing inner
circle voter and tries again.

The initiator discards repairs that disagree with the vote of the
supplier of the repair and removes the supplier from the reference list.
The inconsistency between a vote and the AU on which that vote was
purportedly computed may signal a fault at the repair supplier.

Note that the initiator need only make up to $D$ repair attempts.
If during repairs
the initiator has agreed with
more than $D$ but fewer than $V-D$ voters in total, it knows that
reaching a landslide win through subsequent repairs is impossible.
It deems the poll inconclusive, raising the corresponding alarm
(Section~\ref{sec:protocol:specification:alarms}).

\subsubsection{Reference List Update}
\label{sec:protocol:specification:referenceListUpdate}

Once a poll has concluded successfully, whether initially or after a
repair, the initiator updates its reference list by the following four
steps.  First, it removes those peers on whose votes it based its
decision.  Specifically, it removes all disagreeing inner circle voters
and enough randomly chosen agreeing inner circle voters to make the
total number of removals $Q$ (see
Section~\ref{sec:protocol:analysis:effortExpiration}).  Second,
it resets the \emph{time inserted} field of the remaining
agreeing inner circle voters in the reference list with the current poll
counter.  Third, it inserts all outer circle peers whose votes were
valid and agreeing (with the eventual contents of the AU, after any
potential repairs).  Fourth, it inserts randomly chosen entries copied
from its friends list up to a factor $C$ of the reference list size (the
reference list is \emph{churned} --- see
Section~\ref{sec:protocol:analysis:churning}).  Finally, it removes all
peers that have not voted in the last $E$ polls it has called, i.e.,
those entries whose \emph{time inserted} is at least $E$ polls less than
the current poll counter.

A poll may fail to attract $Q$ or more valid votes from inner circle
participants. If so, the poll initiator ignores all disagreeing votes,
but refreshes or inserts into the reference list the agreeing votes
from both circles.

\subsubsection{Poll Solicitation}
\label{sec:protocol:specification:solicitation}

This and subsequent sections describe the opinion poll protocol from the
point of view of an invitee.

When a LOCKSS peer receives a \msg{Poll} message from a poll initiator,
it chooses a fresh, random \emph{challenge} value, a fresh, random
Diffie-Hellman public key, and computes a symmetric session key from it
and from the poll initiator's public key included in the \msg{Poll}
message.  If the peer is not currently the initiator of, or a voter in,
another poll it decides to vote in this new poll.
It sends back a \msg{PollChallenge} message of the form
[\emph{Poll ID}, \emph{DH Public Key}, \emph{challenge}, \emph{YES}].
Otherwise, it declines to vote and responds with a
\msg{PollChallenge} message of the
form [\emph{Poll ID}, \emph{DH Public Key}, \emph{challenge}, \emph{NO}].
In either case, the
challenge and the YES/NO bit are encrypted with the session key.

Finally, the peer sets an \emph{effort timer} and waits for a
\msg{PollProof} message from the poll initiator (see
Section~\ref{sec:protocol:specification:effort}).  If the message
never arrives, the peer discards all poll state.  Otherwise, the peer
verifies the \msg{PollProof} message.

\subsubsection{Poll Effort Verification}
\label{sec:protocol:specification:effortVerification}

A voter verifies the poll effort proof it receives in a
\msg{PollProof} message using the poll identifier and the challenge it
sent to the initiator
(Section~\ref{sec:protocol:specification:solicitation}).  If the
verification succeeds, the voter chooses $I$ other peers at
random from its own reference list, and nominates them for inclusion
into the poll initiator's outer circle via a \msg{Nominate} message of
the form [\emph{Poll ID}, \emph{Nominations}] encrypted with the session
key.  Then the voter constructs its vote.

\subsubsection{Vote Construction}
\label{sec:protocol:specification:voteConstruction}

A vote consists of a hash of the AU interleaved
with provable computational effort.

Vote computation is divided into rounds, each returning a proof of
computational effort and a hash of this proof with a portion of the AU.
In each round, the computational effort and the AU portion that is
hashed both double in size (see
Section~\ref{sec:protocol:analysis:effortSizing}).
The first round takes as input, and is
dependent upon, the challenge and poll identifier.  Subsequent rounds
take as input, and are dependent upon, the output of the previous round.
The voter sends the proofs of computational effort and AU hashes from
all rounds in a single encrypted \msg{Vote} message to the poll
initiator.

A peer who refused to participate in the poll sends back to the
initiator an encrypted \msg{Vote} message with bogus contents.

\subsubsection{Repair Solicitation}
\label{sec:protocol:specification:repairSolicitation}

After the vote,
the initiator of a poll may request a voter to supply a repair
via a \msg{RepairRequest}
message (Section~\ref{sec:protocol:specification:repair}).  If that
voter has conducted a poll on the same AU in the past, in which
the initiator supplied a valid agreeing vote, then the voter
responds to the request with a \msg{Repair} message.  The \msg{Repair}
message contains the poll identifier and its own copy of the AU,
encrypted with the symmetric session key.  Otherwise, the voter
discards the request.

\subsubsection{Alarms}
\label{sec:protocol:specification:alarms}

LOCKSS peers raise alarms when they suspect that an attack is under
way. The alarm requests human involvement in suppressing the attack, and
is thus expensive.

An \emph{inconclusive poll alarm} suggests that the library should
contact others, examine the differences between their copies and
determine a cause.  Any compromised nodes found during this process
are repaired.  If the institutions hosting the peers voting for
bad copies cannot be identified or do not cooperate, their peers are blacklisted.

A \emph{local spoofing alarm} suggests that the network surrounding the
peer should be audited and any compromised nodes removed.  The cost of
this alarm can be reduced by placing peers on their own subnets.

An \emph{inter-poll interval alarm} is raised if no poll has reached
quorum in several average inter-poll intervals.  An attrition attack may
be underway (Section~\ref{sec:adversary:attacks}), or the peer may no
longer be fast enough to keep up with the system; human attention is
needed in either case.  Logs with large numbers of poll requests from
previously unknown peers might lead to potential attackers who should be
blacklisted.

\subsection{Protocol Analysis}
\label{sec:protocol:analysis}

To defend the LOCKSS system from attack, we make it costly
and time-consuming for an adversary to sway an opinion poll in his favor or
to waste loyal peers' resources.
This means the protocol must
\begin{itemize}
\item prevent the adversary from gaining a
foothold in a poll initiator's reference list (prevent him from
populating it with malign peers),
\item make it expensive
for the adversary to waste another peer's resources, and
\item make it likely that the adversary's attack will be detected before
it progresses
far enough to cause irrecoverable damage.
\end{itemize}
We use provable recent effort, rate limiting,
reference list churning,  and obfuscation of protocol state to make
it expensive and slow for an adversary to gain a significant
foothold in a peer's reference list or waste other peers resources.
We raise alarms when we detect signs of attack.

\subsubsection{Effort Sizing}
\label{sec:protocol:analysis:effortSizing}

One application of our principle of inertia
(Section~\ref{sec:designPrinciples}) is that large changes to the system
require large efforts.  In a protocol where some valid messages cost
nothing to produce, but cause the expenditure of great effort --- e.g.,
a cheap request causing its recipient to hash a large amount of data ---
this principle is unfulfilled.  To satisfy our inertia requirement in
LOCKSS, we adjust the amount of effort involved in message exchanges for
voting, discovery, and poll initiation, by embedding extra, otherwise
unnecessary effort.

For this purpose we need a mechanism satisfying at least three
requirements.  First, it must have an adjustable cost, since different
amounts of additional effort are needed at different protocol steps.
Second, it must produce effort measurable in the same units as
the cost it adjusts (hashing in our case).  Third, the cost of
generating the effort must be greater than the cost of verifying it,
which makes abuse expensive.

We use a mechanism for provable effort based on a class of
\emph{memory-bound functions}~\cite{Abadi2003} (MBF) proposed by Dwork
et al.~\cite{Dwork2003} to prevent email spam.  These cryptographic
functions have a computation phase, yielding a short \emph{proof of
effort}, and a verification phase, which checks the validity of that
proof.  A parameter of the system sets the asymmetry factor by which the
computation phase is more time consuming than the adjustable cost of the
verification phase.  We provide details on how we set the parameters of
the MBF mechanism in Appendix~\ref{sec:economicDesign}.

MBFs are attractive for our purposes because the inherent cost of the
hashing necessary for voting is also memory-bound, and because the
difference in performance between available memory systems is much less
than the difference in other characteristics such as CPU speed~\cite{Douceur2002}.  This
observation has persisted across generations of technology.
Nevertheless, if another mechanism for imposing cost becomes more
attractive, the protocol could easily be revised to use it; it is the
concept of imposing costs on peers that is important rather than the
particular mechanism we use.

In voting, the cost of constructing a vote must be greater than the cost
of processing the vote.  We interleave AU
hashing with effort proof generation and transmit the resulting
proofs in the \msg{Vote} message
(Section~\ref{sec:protocol:specification:voteConstruction}).
This ensures that bogus votes causing the poll initiator to hash
its AU in vain are more expensive to create than the effort
they waste.
The extra effort is interleaved with the hashing \emph{in rounds}
to prevent a cheap, bogus vote from wasting a lot of verification effort.
The rounds ensure that generating a vote that is valid up to round $i-1$ but then
invalid costs more than its verification up to the $i$-th round.

In discovery, we require peers found via others' nominations to
participate first in the outer circle of a poll and generate a valid
agreeing but ineffectual vote before they are invited into the reference
list.  They must thus prove substantial effort (and wait on average more
than $R$) before they are able to affect the result of a poll.  This
makes it expensive and time-consuming for a malign peer to get an
opportunity to vote maliciously and effectively.

Finally, in poll initiation, an initiator must expend more effort than
the cumulative effort it imposes on the voters in its poll.
Otherwise, a malign peer would initiate
spurious polls at no cost, causing loyal peers to waste their resources.
We require the poll initiator to prove more effort in the
\msg{PollEffort} message (Section~\ref{sec:protocol:specification:effort})
than the voter needs to verify that effort and then construct its vote.

\subsubsection{Timeliness Of Effort}
\label{sec:protocol:analysis:effortExpiration}

Our principle of avoiding third-party reputation leads us to
use several techniques to ensure that only proofs of recent
effort can affect the system. They prevent an adversary from exploiting
evidence of good behavior accumulated over time.  The requester of a
poll effort proof supplies a challenge on which the proof must depend;
the first round of a subsequent vote depends on that proof, and each
subsequent round depends on the round preceding it. Neither proofs nor
votes can be precomputed.

Peers must supply a vote, and thus
a proof of effort, to be admitted to the reference list, except for
churning.
If several polls take place without the admitted peer taking part,
perhaps because it died, the initiator removes it.  If the
admitted peer is invited and does take part in a poll, it must
supply a vote and thus further proof of effort.  After the poll
the initiator ``forgets'' the peer if it disagreed,  or if it
agreed but was chosen among the Q peers removed
(Section~\ref{sec:protocol:specification:referenceListUpdate}).
To limit the effect of this removal on loyal peers' reference list size,
we treat the poll as if it had a bare quorum and remove only the
corresponding number of agreeing peers;
additional agreeing peers do not affect the result and are thus treated
as part of the outer circle for this purpose.

By these means we ensure that any peer identity, whether loyal or
malign, must continually be sustained by at least a minimum rate of
expenditure of effort if it is not to disappear from the system.
Although the lack of long-term secrets makes it cheap for an
adversary to create an identity,  sustaining that identity for
long enough to affect the system is expensive.  Unless the adversary's
resources are truly unlimited, there are better uses for them
than maintaining identities that do not contribute to his goals.

\subsubsection{Rate Limiting}
\label{sec:protocol:analysis:rateLimiting}

Another application of our principle of inertia
(Section~\ref{sec:designPrinciples}) is that the system should not
change rapidly no matter how much effort is applied to it.
We use rate-limiting techniques to implement this.

Loyal peers call polls autonomously and infrequently, but often enough
to prevent random undetected damage from affecting readers
significantly.  This sets the effort required of the voters, and means
that an adversary can damage a loyal peer only when that peer calls a
poll.  \emph{The rate at which an attack can make progress is limited by
the smaller of the adversary's efforts and the efforts of his victims.}
The adversary cannot affect this limit on the rate at which he can
damage loyal peers.

\subsubsection{Reference List Churning}
\label{sec:protocol:analysis:churning}

A protocol attacker (Section~\ref{sec:adversary:attacks})
needs to populate a loyal peer's reference list with malign peers
as a precondition for damaging its AU.
We reduce the predictability of the mechanism by which the reference
list is updated using \emph{churning}.

It is important for a peer to avoid depending on a fixed set of
peers for maintenance of its AU, because those peers may become
faulty or subversion targets.
It is equally important not to depend entirely on peers nominated
by other peers of whose motives the peer is unaware.

By churning into the reference list a few peers from its friends list in
addition to the outer circle agreeing voters, the initiator hampers
attempts to fill the reference list with malign conspirators.  Absent an
attack, the proportion of malign peers in both the outer circle and the
friends list matches the population as a whole.  An adversary's attempt
to subvert the random sampling process by nominating only malign peers
raises the proportion of 
malign peers in the outer circle but not in the friends list.  Churning
reduces the effect of the attack because, on average, the friends list is
less malign than the outer circle, \emph{even when the initial friends
list contains subverted peers}.

\subsubsection{Obfuscation of Protocol State}
\label{sec:protocol:analysis:obfuscation}

Our design principles (Section~\ref{sec:designPrinciples})
include assuming a powerful adversary,
capable of observing traffic at many points in the network.
We obfuscate protocol state in two ways to deny him
information about a poll other than that obtained from the
malign participants.

First, we encrypt all but the first protocol
message exchanged by a poll initiator and each potential voter,
using a fresh symmetric key for each poll and voter.
Second, we make all loyal peers invited into a poll,
even those who decline to vote, go through the motions of the
protocol, behind the cover of encryption.  This prevents an adversary from
using traffic analysis to infer state such as the number of loyal peers who
actually vote in a specific poll.  Note
that in our modeled adversary and simulations we conservatively assume that the adversary
\emph{can} infer such information.

\subsubsection{Alarms}
\label{sec:protocol:analysis:alarms}

In accordance with our design principle that intrusion detection be
inherent in the system, the protocol raises an alarm when a peer
determines that a poll is inconclusive, suspects local spoofing, or has
been unable to complete a poll for a long time.  Raising an
alarm is thus expensive; a significant rate of false alarms would render
the system useless (Section~\ref{sec:rateOfFalsePositives}).

The expectation is that alarms result in enough loss
to the adversary,
for example by causing operators to remove damage,
malign peers and compromised nodes,
that a rational adversary will be highly motivated to avoid them, unless
raising alarms is his primary goal.

\section{Adversary Analysis}
\label{sec:adversary}

A peer-to-peer system running on a public network must expect to be
attacked, even if the attackers have nothing tangible to gain.  We
present the capabilities we assume of adversaries
(Section~\ref{sec:adversary:capabilities}), explore the space of attacks
they can mount in terms of goals (Section~\ref{sec:adversary:attacks}),
and identify specific attack techniques available to adversaries
(Section~\ref{sec:adversary:techniques}).
Finally,
we describe in detail the particular adversaries we study in this paper
(Section~\ref{sec:adversary:strategies}).

\subsection{Adversary Capabilities}
\label{sec:adversary:capabilities}

The adversary we study in this paper controls a group of malign peers.
We believe the following abilities match our ``powerful adversary''
design principle:
\begin{description}
\item[Total Information Awareness] Malign peers know each other.  Any
information known to one malign peer, including the identities of other
malign peers involved in a poll, is immediately known to all.

\item[Perfect Work Balancing] Any of the adversary's nodes can perform
work on his behalf and relay it instantaneously to the node presumed
to have performed it.

\item[Perfect Digital Preservation] The malign peers have magically
incorruptible copies of both the good AU and as many bad ones as they
require.

\item[Local Eavesdropping] The adversary is aware of the existence and
contents of packets originating or terminating at a network on which
he controls a physical node.  He cannot observe
the existence or contents of packets that originate and terminate
anywhere else.

\item[Local Spoofing] From a routing realm in which he controls a
physical node, the adversary can send IP packets whose ostensible
source is any local address and destination is any Internet address;
or he can send packets whose ostensible source is any Internet address
and destination is a local address.

However, the adversary cannot usefully send IP packets whose ostensible
source and destination addresses are from routing realms within which he
has no control.  The protocol's encryption handshake prevents peers
from taking part in polls unless they can receive packets at their ostensible
address (see Section~\ref{sec:protocol:specification:solicitation}).
The adversary cannot benefit from spoofing IP addresses on whose traffic
he cannot eavesdrop.

\item[Stealth] A loyal peer cannot detect that another peer executing
the LOCKSS protocol is malign.

\item[Unconstrained Identities] The adversary can increase the number of
identities he can assume in the system, by
purchasing or spoofing IP addresses.

\item[Exploitation of Common Peer Vulnerabilities] The adversary can
instantaneously take over those peers from the LOCKSS population that
run an implementation with the same exploitable vulnerability.  He can
then instantly change the state and/or the logic of the afflicted peers,
causing them to become malign.

\item[Complete Parameter Knowledge] Although loyal peers actively
obfuscate protocol state
(Section~\ref{sec:protocol:analysis:obfuscation}), we assume that the
adversary knows the values of protocol parameters, including $Q$, $N$,
$A$ and $D$, as set by loyal peers.
\end{description}
We measure the adversary by the extent to which he can subvert loyal
peers by any means, and by the total computational power he has
available.  Our analyses and simulations do not depend on identifying
the cause of an individual peer becoming malign.

\subsection{Adversary Attacks}
\label{sec:adversary:attacks}

Attacks against LOCKSS can be divided according to their functional
approach into \emph{platform} attacks and \emph{protocol} attacks.
Platform attacks seek to subvert the system hosting the LOCKSS
implementation, either to tamper with the logic and data of LOCKSS
from the inside or to use a LOCKSS peer as a jumping point for further
unrelated attacks.  Our implementation makes considerable efforts to
resist platform attacks~\cite{Rosenthal2003b}.  We do not address
platform attacks further in this paper.  We do, however, allow our
simulated adversary to mount such attacks successfully, taking over a
substantial portion of the peer population at the beginning of the
simulations (See Section~\ref{sec:simulationEnvironment}).

We can divide protocol attacks according to the role within
the LOCKSS protocol whose operations the adversary manipulates for his
purposes.  In general, inner circle peers within a poll command greater
power for mischief than outer circle peers, poll initiators can hurt
their own polls but not much more, and spoofing and eavesdropping are
most helpful when available near a loyal poll initiator's network.  We
analyse attacks by role in Section~\ref{sec:rollAttacks}.

Finally, we can divide attacks according to the goals
of the adversary.  Different adversary goals require different
combinations of platform and protocol attacks.  We list brief
descriptions of some possible adversary goals below:
\begin{description}
\item[Stealth Modification] The adversary wishes to replace the
  protected content with his version (the \emph{bad} content).  His goal
  is to change, through protocol exchanges, as many replicas of the
  content held by loyal peers as possible without being detected, i.e.,
  before the system raises an alarm.  Measures of his success are
  the proportion of loyal peers in the system hosting replicas of the
  bad content at the time of the first alarm, and the probability that a
  reader requesting the content from any peer in the system obtains the
  bad content.
\item[Nuisance] The adversary wishes to raise frequent, spurious
  LOCKSS alarms to dilute the credibility of alarms and to waste
  (primarily human) resources at loyal peers.  A measure of adversary
  success is the time it takes for the adversary to raise the first
  alarm (the lower, the better for the adversary).
\item[Attrition] The adversary wishes to prevent loyal peers from
  repairing damage to their replicas caused by naturally occurring
  failures.  Towards that goal, he wastes the computational resources
  of loyal peers so that they cannot successfully call polls to audit
  and repair their replicas.  Measures of his success are the
  time between successful polls called by loyal peers or the busyness
  of loyal peers (in both cases the higher, the better for the adversary).
\item[Theft] The adversary wishes to obtain published content without
  the consent of the publisher.  For example, he wishes to obtain
  for-fee content without paying the fee.  A measure of adversary
  success is the time to obtain a copy of the restricted content (the
  lower, the better for the adversary).
\item[Free-loading] The adversary wishes to obtain services without
  supplying services to other peers in return.  For example, he wishes
  to obtain repairs for his own replicas, without supplying repairs to
  those who request them from him.  A measure of his success is the
  ratio of repairs supplied to repairs received (the lower, the better
  for the adversary).
\end{description}

In this paper we focus on adversaries with the stealth modification,
nuisance, and attrition goals.  We choose these three goals because they
have the greatest potential, especially in concert, to disrupt the
preservation work of a community of LOCKSS peers, since they can
directly or indirectly hurt the preserved content \emph{and} discredit
the functions of the system that can detect the damage.  In the rest of
this section, we briefly describe what LOCKSS can do about theft and
free-loading.

The LOCKSS system does not materially increase the risk of theft.
Repairs are the only protocol exchanges in which content is
transferred.  A peer supplies a repair only if the requester has
previously proved with an agreeing vote that it once had the same
content (see
Section~\ref{sec:protocol:specification:repairSolicitation}).  The
protocol cannot be used to obtain a first instance of an AU.  Most
e-journals authorize ranges of IP address for institutional
subscribers; existing LOCKSS peers use this mechanism to authorize
their crawls for first instances of content.  If more secure forms of
content authorization (e.g., Shibboleth~\cite{Shibboleth}) become
widely accepted, LOCKSS peers can use them.

The repair mechanism also limits the problem of free-loading.
The primary good exchanged in LOCKSS, content repair,
is only available to peers who prove through voting that they have had
the content in the past.  A peer who votes dutifully --- so as to be
able to obtain repairs if necessary --- but does not itself supply
repairs to others can hardly be described as \emph{free}-loading, given
the cost of the voting required.  Tit-for-tat refusal to supply
repairs to such antisocial peers might further deter this behavior.

In the next section, we present attack techniques available to the
adversary.  We combine these techniques in later sections, to produce
adversary strategies towards the stealth modification, nuisance, and
attrition goals.

\subsection{Attack Techniques}
\label{sec:adversary:techniques}
In this section, we identify possible attack vectors against the LOCKSS
opinion poll protocol, and we describe how they can be exploited.

\subsubsection{Adversary Foothold in a Reference List}
\label{sec:adversary:techniques:referenceList}
The composition of a loyal peer's reference list is of primary
importance.  A loyal peer chooses inner circle participants for a poll
that it initiates from its reference list at random (see
Section~\ref{sec:protocol:specification:initiation}); as a result, the
proportion of malign peers in the inner circle approximates
their proportion in the poll initiator's reference list.
An adversary wishing to control the outcome of a
poll initiated by a loyal peer can increase his chances 
by increasing the proportion of malign peers in
that peer's reference list.

To gain an initial foothold in loyal peers' reference lists, the
adversary must take over peers that used to be loyal.  He does this,
for example, by exploiting common implementation vulnerabilities, 
or by coercing peer operators to act as he wishes.
He can then increase that foothold, but to do so he must wait until a
loyal peer invites a malign peer into the inner circle of a poll.
When invited,  the adversary causes the malign peer to nominate other
malign peers unknown to the poll initiator.  Note that loyal peers also
inadvertently nominate malign peers.

Malign peers in loyal peers' reference lists must behave as loyal
peers until an attack requires otherwise.  Each such peer thus
consumes adversary resources as it must both vote in and call polls
to avoid detection and maintain its position in the list
(Section~\ref{sec:protocol:analysis:effortExpiration}).

We measure the adversary's success at maintaining a foothold
in loyal peer's reference lists with the average \emph{foothold ratio}
over the population of loyal peers.
For a given reference list, the foothold ratio is the proportion
of the list occupied by malign peers.

The LOCKSS protocol has two lines of defense against reference list takeover.
First, loyal peers only change their reference lists after a
poll that they call; the adversary must wait until they do so
before he can increase his foothold (see Section~\ref{sec:protocol:analysis:rateLimiting}).
Second, churning of the reference list allows
the operator of a loyal peer to trade off the risks of depending too
much on a static set of friendly peers against those of depending too
much on peers nominated by others for the outer circle of polls (Section~\ref{sec:protocol:analysis:churning}).

\subsubsection{Delayed Commitment}
\label{sec:adversary:techniques:delayedCommitment}

The adversary need not decide in advance how each of its malign peers
will react to particular poll invitations.  Instead he can determine how
a particular malign peer behaves after having collected all available
information about a poll or its initiator, as per his Total Information
Awareness and Stealth capabilities
(Section~\ref{sec:adversary:capabilities}).

Loyal peers can defend against this adaptive adversary
by requesting commitments on future protocol steps as early as possible.
During repairs, the requester checks that the repair it
receives is consistent with the vote of the repair supplier.
During voting, the poll initiator could request from potential voters
an early commitment on the hash of a few bytes of the AU, chosen randomly.
Later, the poll initiator could verify that each vote is consistent
with the AU version to which that voter committed.
This would reduce the adversary's ability to attack only polls he is
sure of winning and increase the probability of detection.
Our simulations currently use the former defense but not the latter.

\subsubsection{Peer Profiling}
\label{sec:adversary:techniques:peerProfiling}

Using the Local Eavesdropping capability, the adversary can
observe a loyal peer's traffic and attempt to infer useful
information such as likely members of the peer's
reference list, likely participants in a poll, and whether or not the
peer has agreed to vote in a poll.  The adversary can use this
information to make better decisions, as with delayed commitment,
or to mount flooding attacks against invited
loyal peers whose sessions he can hijack
(Section~\ref{sec:adversary:techniques:sessionHijacking}).

Loyal peers defend against eavesdroppers by actively obfuscating their
protocol state (Section~\ref{sec:protocol:analysis:obfuscation}).

\subsubsection{Session Hijacking}
\label{sec:adversary:techniques:sessionHijacking}

The LOCKSS system lacks stable identities because it cannot support
them without long-term secrets.
An attacker can thus impersonate loyal peers and hijack their sessions,
using his Local Spoofing capability (Section~\ref{sec:adversary:capabilities}).
By spoofing local source addresses in messages it sends to remote peers,
or remote source addresses in messages it sends to local peers,
a malign peer within spoofing range of a poll initiator can affect
the poll result by either hijacking sessions between the poll initiator
and loyal invitees, or discrediting loyal invitees of his choosing.

The malign peer can hijack a session by responding to the initiator's
\msg{Poll} message with a spoofed \msg{PollChallenge} message
establishing a session key.  If the initiator also receives the
genuine \msg{PollChallenge} message, the two conflict, the
invitee is discredited,  and the hijack  fails.
If, however, the loyal invitee fails to respond with a \msg{PollChallenge}
or the adversary manages to suppress it, perhaps by flooding the
loyal invitee's link, the hijack can succeed and the malign
peer can vote in the loyal invitee's place.
Once established,
a session is protected by a session key and can no longer be hijacked.

Alternatively, the malign peer local to the poll initiator can
selectively discredit loyal invitees by also responding with a spoofed, conflicting
\msg{PollChallenge}.
Votes based on either challenge are not tallied, as the
initiator lacks a means to distinguish between them.
By suppressing loyal votes in this way the adversary can
increase his foothold ratio
(Section~\ref{sec:adversary:techniques:referenceList}),
or waste human resources by raising spoofing alarms
(Section~\ref{sec:protocol:analysis:alarms}).

The operators of a loyal peer can defend against hijacking by
checking its network vicinity for subverted peers or routers,
by providing it with a dedicated subnet,
and by monitoring for packets spoofing the router's MAC address.
Loyal peers could retransmit their potentially suppressed
\msg{PollChallenge} messages at random intervals throughout the
poll.  If any of these retransmissions get to the initiator, the
hijacked session is discredited.  This would force the
adversary to suppress traffic from the hijacked peer for
many hours, increasing the probability of detection.

\subsection{Attack Strategies}
\label{sec:adversary:strategies}

In this section, we take a closer look at three kinds of adversaries:
first, an adversary whose goal is to modify an AU stealthily across
loyal peers so as to change the scientific record persistently
(Section~\ref{sec:adversary:strategies:stealth}); second, an adversary
whose goal is to discredit the LOCKSS alarm mechanism, thereby rendering
it incapable of reacting to attacks
(Section~\ref{sec:adversary:strategies:nuisance}); and third, an
adversary whose goal is to slow down LOCKSS long enough for random
faults to cause irrecoverable damage in the content.  For all three
types of adversaries, we describe an attack strategy.  In
Section~\ref{sec:simulation}, we measure through simulation how LOCKSS
fares against such attacks.

\subsubsection{Stealth Modification Strategy}
\label{sec:adversary:strategies:stealth}
The stealth adversary has to balance two goals: changing the consensus
on the target AU and remaining undetected.  To achieve these goals, he
must repeatedly find a healthy poll initiator, convince it
that it has a damaged AU replica without causing it to raise any alarms,
and then, if asked, conveniently offer it a repair with the bad version of
the AU.  This strategy relies primarily on building a foothold in loyal
peers' reference lists
(Section~\ref{sec:adversary:techniques:referenceList}) and on delayed
commitment (Section~\ref{sec:adversary:techniques:delayedCommitment}).

The stealth adversary acts in two phases.
First he \emph{lurks} seeking to build a foothold in loyal
peers' reference lists but otherwise behaving as a loyal peer,
voting and repairing with the correct version of the AU.
Then he \emph{attacks}, causing his malign peers to vote and
repair using either the correct or the bad version of the AU, as needed.
During the attack phase malign peers vote with the correct copy
unless a poll is \emph{vulnerable}, i.e., one in which the overwhelming
majority of the inner circle is malign.  In vulnerable polls malign
peers vote with the bad copy,  because by doing so they can change
the loyal initiator's AU without detection.  Polls are vulnerable if
the following three conditions hold:
\begin{eqnarray}
M + L &\geq& Q \label{eqn:moreThanQuorum}\\
M     &>   & L \label{eqn:majority}\\
L     &\leq& D \label{eqn:landslide}
\end{eqnarray}
Condition~\ref{eqn:moreThanQuorum} ensures that the $V = M + L$ peers
agreeing to vote satisfy the quorum $Q$.
Condition~\ref{eqn:majority} ensures that the $M$ malign peers
determine the result with an absolute majority of the votes.
Condition~\ref{eqn:landslide} ensures that the $L$ loyal peers
are not enough to raise an inconclusive poll alarm at the
initiator.
Our modeled adversary has Complete Parameter Knowledge
(Section~\ref{sec:adversary:capabilities})
and can evaluate this vulnerability criterion exactly, in accordance to
our  ``strong adversary'' design principle.
In a practical system an adversary would have only estimates of
$L$, $Q$, and $D$,
and would thus run a higher risk of detection than in our simulations.

The protocol provides several defenses that are especially relevant
against the stealth adversary.  Individually none is very strong; in
combination they are quite effective. First, in accord with our
rate-limiting principle, the adversary cannot induce loyal peers to call
vulnerable polls but has to wait until they occur.

Second,  a damaged peer continues to call and vote in polls
using its now bad copy of the AU.  Unlike malign peers, it does not
evaluate the vulnerability criterion or decide between the good and
bad versions.  If more than $D$ damaged peers take part in a
poll called by a healthy peer but the adversary deems the poll
invulnerable, an inconclusive poll alarm is raised.

Third, a damaged peer continues to call polls and may invite enough
healthy peers into its inner circle to repair the damage.  For each
loyal peer the stealth adversary damages,  he must expend resources
to maintain his foothold in the peer's reference list and vote
whenever it invites malign peers until the bad version of the AU
prevails everywhere.

Finally, if the stealth adversary fools the initiator of a vulnerable poll
into requesting a repair, he must ensure that the request will go
to one of the malign peers.  The initiator requests repairs
only from peers in whose polls it has voted; others would refuse
the request as they lack evidence that the requester once had a valid copy
(Section~\ref{sec:protocol:specification:repairSolicitation}).
Thus the stealth adversary must expend effort to call polls as well as
vote in polls called by the loyal peers.

\subsubsection{Nuisance Strategy}
\label{sec:adversary:strategies:nuisance}
The nuisance adversary has a simple goal: to raise alarms at loyal peers
as fast as possible.  The nuisance adversary can cause all three types
of alarms (Section~\ref{sec:protocol:analysis:alarms}).

To cause an inconclusive poll alarm, the adversary can use delayed
commitment.  Every time some of his malign peers are invited into a
poll, he evaluates the vulnerability criterion
\begin{eqnarray}
M + L &\geq& Q \label{eqn:moreThanQuorum1}\\
L     &>   & D \label{eqn:moreGoodThanDissent}\\
M     &>   & D \label{eqn:moreBadThanDissent}
\end{eqnarray}
Apart from reaching a quorum (Condition~\ref{eqn:moreThanQuorum1}), the
criterion means that neither are the loyal votes few enough to lose
quietly (Condition~\ref{eqn:moreGoodThanDissent}), nor are the malign
votes few enough to allow a landslide win by the loyal votes
(Condition~\ref{eqn:moreBadThanDissent}).  When the adversary detects
that this criterion is met, he instructs his malign peers to vote in the
poll with a bogus AU, thereby causing the poll to raise an inconclusive
alarm.  If the criterion is not met, the adversary instructs his malign
peers to vote with the correct AU.  This is necessary, so that malign
peers can remain in the reference list of the poll initiator (see
Section~\ref{sec:protocol:specification:tabulation}) for another try in
the future; the adversary can foster the satisfaction of the
vulnerability criterion by building a foothold in the reference list of
the poll initiator
(Section~\ref{sec:adversary:techniques:referenceList}).

The adversary can cause a spoofing alarm by sending conflicting
\msg{PollChallege} messages to the poll initiator.  This strategy can
yield spoofing alarms when the adversary has more than $A$ malign peers
in the invited inner circle of a poll.  Then, without having a spoofer
near the poll initiator, the adversary can cause the initiator to
suspect a local spoofer and to raise frivolous alarms, which waste human
resources at the initiator.

The adversary can cause an inter-poll interval alarm by foiling loyal
peers' attempts to call a poll successfully for long enough.  This
strategy is similar to a strategy with the goal of attrition
(Section~\ref{sec:adversary:strategies:attrition}).

The nuisance adversary need not have his malign peers call polls,
because he does not seek to obtain or supply repairs.  However, he needs
to have his malign peers invited in polls.  As a result, he must follow
the observable aspects protocol (voting with valid votes when asked)
unless he can attack a particular poll.

LOCKSS defends against the nuisance adversary primarily via the autonomy
with which loyal peers decide when to call polls.  As a result, a
nuisance adversary must persist for a while, increasing his foothold in
the reference lists of loyal peers, before he can attack a particular
poll raising an inconclusive poll alarm.

\subsubsection{Attrition Strategy}
\label{sec:adversary:strategies:attrition}

The attrition adversary's goal is to occupy the time the loyal peers
have available for voting,  making it less likely that a poll called by
a loyal peer will gain a quorum of voters.  Success is measured
by the average time between successful, quorate polls at loyal peers.
If the attrition adversary can increse this interval enough,  random
damage at the loyal peers can accumulate and degrade the system.

The attrition adversary's strategy is to call polls as fast as
possible,  inviting only loyal peers.  The adversary's peers do not
vote in polls called by other peers; there is no need to persuade
loyal peers to fetch repairs from them.  We do not yet use a
``newcomer pays extra'' strategy so the attrition adversary
can currently use one-time throw-away identities to call the polls.

The attrition adversary's impact on the system is currently limited
only by the rate at which he can compute the proofs of effort
demanded by the loyal voters he is trying to involve in his polls.
We are investigating techniques that limit his impact more
effectively (see Section~\ref{sec:future}).

\subsection{\label{sec:rollAttacks}Attack Anaylsis by Peer Role}
We divide the description of the attacks according to the role played
in the poll by the malign peer.
For each role,
we examine attacks that affect the result of the poll,
and those that degrade the quality of the initiator's reference list.

We also describe additional attacks that deny service,
and some spoofing attacks.
\subsubsection{\label{sec:pollInitiator}Poll initiator}
The worst a malign poll initiator can do is to deny service.

\begin{description}
\item[Affect poll result] A malign poll initiator can affect the result
of a poll he initiates, but to no effect on the rest of the system,
since he is already malign.

\item[Degrade reference list] A malign poll initiator can degrade his
own reference list, but to no effect on the rest of the system, since he
is already malign.

\item[Deny service] The attrition adversary acts as a malign poll
initiator and invites large numbers of loyal peers into polls to waste
their resources and prevent them from coming to consensus.  We set the
poll effort cost to make this attack very expensive
(Section~\ref{sec:protocol:analysis:effortSizing}), and raise an
inter-poll interval alarm if we detect it.
\end{description}

\subsubsection{\label{sec:innerCircle}Inner circle}

A malign peer invited into the inner circle of a loyal peer's poll can
take part in all three kinds of attacks.

\begin{description}
\item[\label{sec:VulnerablePoll}Affect poll result] If enough malign 
peers are invited into a poll they can affect the
result without detection (Section~\ref{sec:adversary:strategies:stealth}).
The nuisance adversary attacks a poll if it meets the criteria discussed
in Section~\ref{sec:adversary:strategies:nuisance}).

\item[Degrade reference list] A malign inner circle invitee has the
opportunity to degrade the quality of the initiator's reference list, by
recommending other malign peers into the poll's outer circle.  Malign
peers recommended for the outer circle will only get into the loyal poll
initiator's reference list if they expend effort and vote with the
consensus.

However, the malign peers have an advantage over the loyal peers.  They
know the identities of the other malign peers.  The loyal peers will
recommend $B$ malign peers on average, but the malign peers will
recommend only malign peers.  Doing so costs the adversary effort over
time because the malign invitees have to exert effort to get in and stay
in the loyal peer's reference list.  The loyal peers also have to exert
effort to stay in the malign peer's lists, even if this particular
effort is wasted.  The malign peers do not have to exert effort to stay
in their co-conspirator's lists.

\item[Deny service] A malign inner circle peer can do four 
things to deny service,
only the last of which is effective:

\begin{itemize}
\item It could generate a bogus vote in a poll the malign peers
are certain to lose.
The initiator will eliminate the vote with much less effort than it took
to create and remove the malign peer from its reference list.
This is not a good strategy.

\item It could recommend loyal peers into the outer circle
then discredit the poll
by inviting the same loyal peers into a bogus poll with the same ID.
The effort of creating the bogus poll is larger than the cost
to these loyal
peers of detecting that it is bogus.
The adversary would benefit more from using the effort to degrade the
initiator's reference list further.

\item It could refuse to send a vote, raising the probability that the
poll would fail to get a quorum, at the risk of losing its slot in the
initiator's reference list if the poll does achieve a quorum.  Unless
the adversary is resource-constrained, the malign peer should vote and
recommend other malign peers.

\end{itemize}
\end{description}

\subsubsection{\label{sec:outerCircle}Outer circle}

A malign peer invited into the outer circle of a poll is offered a
chance to occupy a place in the loyal peer initiator's reference list.
This is a valuable opportunity to attack a future vulnerable poll.

\begin{description}
\item[Affect poll result] A malign outer circle invitee cannot change
the result of the poll because the initiator will not consider its vote.

\item[Degrade reference list] To degrade the reference list,
the malign outer circle invitee must vote with the consensus.
It must then continue to act as a healthy peer in future polls
until the malign peers decide to attack.

\item[Deny service] A malign outer circle invitee could also remain mute,
causing a shortage of outer circle participants,
but the adversary would benefit more from degrading the initiator's
reference list.
\end{description}

\section{Simulation}
\label{sec:simulation}
We have evaluated our new protocol's resistance to random failures
and malicious attacks using the simulation we present in this
section. We first describe our simulation environment in
Section~\ref{sec:simulationEnvironment}.  Then we explain how we
simulate loyal peers (Section~\ref{sec:loyalSimulation}).  Then
in Section~\ref{sec:adversarySimulation} we describe how we simulate
the different  adversary strategies from
Section~\ref{sec:adversary:strategies}.
Section~\ref{sec:results} collects our simulation results.

\subsection{Simulation Environment}
\label{sec:simulationEnvironment}
In this section, we describe the simulation environment we use for our
evaluation.  This includes our simulator, the network model we employ,
and our application-layer overlay topology initialization.

We use Narses, a Java-based discrete-event simulator~\cite{Giuli2002}
designed for scalability over large numbers of nodes, large amounts of
traffic, and long periods of time.  Narses offers facilities for a
variety of flow-based network models allowing trade-offs between speed
and accuracy.  The simulator can also model expensive
computations, such as hashes and proofs of effort, allowing some realism
in our simulation of protocols involving cryptographic primitives.

Since we simulate a LOCKSS network for up to 30 (simulated) years, we
use a faster-to-simulate network model that considers
propagation latency but not traffic congestion.  We simulate the
underlying network topology as a star at the center of which lies the
``core.''  Individual nodes are linked to the core via a link whose
bandwidth is chosen at random among 1.5, 10 and 1000 Mbps, and
whose propagation latency is chosen uniformly at random from 1 to 30 ms.
The core has infinite switching capacity; as a result, the effective
bandwidth of a flow from node $A$ to node $B$ is the minimum bandwidth
of the two links,
and its propagation latency is
the sum of the propagation latencies of the two links.

Every simulation run starts with an initial population of 1000 peers,
each storing an AU that takes 120 seconds to hash.  The dynamic contents of the
reference lists of these peers determine the application-layer topology
of the LOCKSS overlay.  As the protocol requires, the reference list of each
peer is initialized with the content of its friends list.
We initialize each peer's friends list with a clustering technique.  
Peers are randomly assigned to clusters of 30 peers.  For each peer, we
add 29 other peers to its friends list, 80\% of which are chosen randomly from
its own cluster and the rest chosen randomly from other clusters.

We simulate a provable effort mechanism similar to the MBF scheme
devised by Dwork et al.~\cite{Dwork2003}. In keeping with the
constraints placed by that scheme and with the requirements we set out
in Section~\ref{sec:protocol:analysis:effortSizing}, we derive one
possible set of provable effort sizes for the protocol
(Appendix~\ref{sec:economicDesign}).  Given that hashing the AU costs
$S$, the poll effort construction size
(Section~\ref{sec:protocol:specification:effort}) is $(20/3)S$, the
verification of a poll effort proof
(Section~\ref{sec:protocol:specification:effortVerification}) costs
$(5/3)S$, the cost of computing a valid vote
(Section~\ref{sec:protocol:specification:voteConstruction}) is $5S$, and
the
cost of verifying a vote 
(Section~\ref{sec:protocol:specification:voteVerification}) is $2S$ for
agreeing and $S$ for disagreeing votes.

If the cost of hashing the AU is 120 seconds, the initiator spends 800
seconds per invitee generating the \msg{PollProof} message and 240
seconds per invitee verifying an agreeing \msg{Vote} message.  Each
invitee spends 200 seconds verifying the \msg{PollProof} message and 600
seconds generating the \msg{Vote} message.  An entire successfully
concluded poll without repairs costs the initiator 1040 seconds of
computation per invitee. With 20 invitees it would take 6 hours, which
is comparable to the duration of polls in the current test.

\subsection{Simulated Loyal LOCKSS Peers}
\label{sec:loyalSimulation}

We simulate loyal LOCKSS peers as simple state machines implementing the
protocol of Section~\ref{sec:protocol}.  We set the protocol parameters
(see Section~\ref{sec:protocol:specification}) to values reflecting
those in the 60-peer tests of the existing system ($N = 20$, $Q
= 10$, $D = 3$, $A = 3$, $I = 10$, $E = 4~\mathit{polls}$), except $R = 3~\mathit{months}$ which
we estimate reflects production use.

We set protocol timers to be just long enough for the slowest machine to
complete the corresponding protocol step.  Peers always consider
themselves the fastest.  For example, peers who have been invited into a
poll give the poll initiator enough time to compute the poll effort
proof for $N$ invitees (see
Section~\ref{sec:protocol:specification:solicitation}), assuming that
the poll initiator has a memory system 5 times slower than
theirs~\cite{Dwork2003}.

Our simulated peers commit to a poll exclusively, for the duration of
that poll, even when idly waiting for a protocol message to come back.
However, a peer that wishes to call its own poll but is also invited in
another poll called by someone else prefers to call its own poll.

All loyal peers in a simulation run have the same nominal rate of random
undetected errors that unobtrusively replace the victim's AU replica
with random bits.

\subsection{Simulated Adversary}
\label{sec:adversarySimulation}
In this section we address our simulation of LOCKSS adversaries.  We
outline how we represent an adversary and his malign peers, and then we
describe how we implement within our simulation environment the attack
techniques available to him (see
Section~\ref{sec:adversary:techniques}).

We simulate an adversary as a multi-homed node with as many network
interfaces (NICs) as the number of IP addresses,
and as many CPUs as the number of nodes controlled by the
adversary (i.e., one \emph{humongous} computer).
The numbers of NICs and CPUs are
parameters of the simulation.
An adversary with few NICs and many CPUs
has a lot of processing power at his disposal, but is without a great
presence in the network.  An adversary with many NICs and
fewer CPUs has some processing power but a lot of scope
for spoofing IP addresses.

To gain a foothold in loyal peers' initial reference lists
(Section~\ref{sec:adversary:techniques:referenceList}), the adversary
may use his ability to take over some of the LOCKSS peer population (see
Section~\ref{sec:adversary:capabilities}).  We initialize simulation
runs at the instant when the take-over is complete.  For example, to run
a simulation where the adversary subverts 30\% of the 1000 peers, but
also has 100 extra CPUs at his disposal, we simulate an adversary with
$1000 \times 30\% + 100 = 400$ CPUs and only 700 loyal peers.

In our simulations, once the
adversary receives a \msg{PollProof} message via one of its NICs,
he considers the number of those NICs via which he has received
\msg{Poll} messages thus far to be $M$ for this poll (see
Section~\ref{sec:protocol:specification}).  Then, the adversary divides
the list of its own NICs among the $M$ malign peers as which he
participates in the inner circle of the poll.  When a particular malign peer NIC
receives a \msg{PollProof} message, the adversary waits the appropriate
time for the verification of the poll effort proof and then responds
with a \msg{Nominate} message holding the corresponding portion of the
list of malign NICs.  The adversary thus ensures that the loyal
poll initiator will insert into its outer circle the maximum number of
malign peer addresses; the adversary cannot do better, without knowledge
of the nominations of other loyal inner circle peers.

We simulate delayed commitment
(Section~\ref{sec:adversary:techniques:delayedCommitment}) by waiting
until the adversary must start computing his first vote before deciding
on which AU that first vote will be.  At that time, the adversary
evaluates the appropriate poll vulnerability criterion, according to the
strategy we simulate, decides whether to attack the poll and how, and
commits to the appropriate version of the AU.

We simulate the adversary's ability to profile loyal peers
(Section~\ref{sec:adversary:techniques:peerProfiling}) by making all
variable protocol parameters known to him.  We do not, in this paper,
otherwise simulate the presence of eavesdroppers near loyal peers.

Finally, we do not simulate in this work the adversary's ability to
hijack poll sessions between loyal poll initiators and their loyal
invitees.

\subsubsection{Simulated Stealth Modification Adversary}
\label{sec:stealthSimulation}

We simulate the effects on LOCKSS of an attack by an adversary
following the stealth modification strategy
(Section~\ref{sec:adversary:strategies:stealth}) in two sets:
\emph{lurking} and \emph{attacking} simulations,
corresponding to the lurking and attack phases of the strategy.  In
lurking simulations, the adversary seeks only to extend his foothold
in loyal peers' reference lists.  After initially subverting some of
the loyal peers, the adversary has malign peers behave exactly as
loyal peers do, except for formulating their \msg{Nominate} messages
as described in Section~\ref{sec:adversarySimulation} above.  Lurking
simulations last 20 simulated years.

In attacking simulations, malign peers seek not only to extend their
foothold in loyal peers' reference lists, but also to change the loyal
peers' replicas of the AU with the bad version that the adversary
wishes to install throughout the community.  Therefore, malign peers
also evaluate the vulnerability criterion and decide, as described in
the previous section, on which AU to base their votes and their repairs.
We initialize the population in an attacking simulation as if a lurking
phase preceded the simulation: we initialize the reference lists of
loyal peers with a given foothold ratio.  Attacking simulations last 10
simulated years, unless an inconclusive poll alarm is raised.

To draw conclusions about entire stealth modification attacks, we must
combine the results of a lurking simulation with the results of a
\emph{compatible} attacking simulation.  We accomplish this by first
running a set of lurking simulations for the set of input parameters we
seek to study.  Based on these runs, we identify how great a foothold
ratio the adversary can obtain, for given input parameters.  Then we run
a set of attacking simulations with input parameters that match the
input parameters of the lurking simulations as well as the observed
possible foothold ratios gained by the adversary.  For example, when
studying the stealth modification adversary who begins by subverting
20\% of the 1000 initially loyal peers, we run a number of lurking
simulations (for different random seeds), from which we conclude that
the adversary can obtain average foothold ratios of 40 to 55\%.  Based
on this, we only run attacking simulations for 20\% subversion of the
1000 initially loyal peers and initial attack-phase foothold ratios that range
between 40 and 55\%.

Splitting the strategy into two sets of simulations allows us to explore
the choice the adversary makes about the foothold he must achieve before
switching from lurking to attacking.  In our results, we
assign the first possible time at which this foothold is achieved for a
given initial subversion as the duration of the lurk phase for that
subversion.

In both lurking and attacking simulations, the adversary calls polls as
a loyal peer would, with two differences.  First, malign peers never
verify votes that they receive on the polls they initiate, since the
adversary does not care about the outcome of the poll.  Second, the
adversary never invites its malign peers into his own polls, since he
calls polls only to convince loyal peers to ask him for repairs.

\subsubsection{Simulated Nuisance Adversary}
\label{sec:nuisanceSimulation}

To simulate the effects on LOCKSS of an attack by an adversary following
the nuisance strategy (Section~\ref{sec:adversary:strategies:nuisance})
we run simulations similar to the attacking simulations of the stealth
modification adversary (Section~\ref{sec:stealthSimulation}).

Nuisance simulations differ from attacking stealth simulations in three
ways.  First, the adversary does not call polls, because he has no
vested interest in maintaining a convincing fa\c{c}ade of compliance.
Second,
he does not lurk but rather attacks any vulnerable polls immediately.
Third,
he uses a weaker vulnerability criterion.

\subsubsection{Simulated Attrition Adversary}
\label{sec:simulation:attrition}
 
We simulate the attrition adversary with unlimited identities but
limited resources.  He calls useless polls to consume the loyal peers'
resources.  Unlike the stealth adversary, he neither lurks to degrade
reference lists nor attacks polls.  Loyal peers give priority to calling
their own polls, their remaining resources go to participating in polls
called by others.  A loyal peer raises an inter-poll interval alarm if
it has not completed a poll in 3 times the expected inter-poll interval.

\section{Results}
\label{sec:results}
In this section we evaluate the new LOCKSS opinion poll protocol through
simulation.  We explore how the protocol deals with random storage
faults, as well as attacks by the stealth modification adversary
(Section~\ref{sec:adversary:attacks}).  We demonstrate the following
points:
\begin{itemize}
\item Absent an attack, substantial rates of random damage at peers
result in low rates of false alarms
(Section~\ref{sec:rateOfFalsePositives}).
\item With up to 1/3 of the peers subverted, the stealth adversary
fails.  Above that, the probability of irrecoverable damage increases
gradually (Section~\ref{sec:results:stealthAdversary}).
\item A nuisance adversary whose goal is simply to raise an alarm has to
exert significant effort over a long period
(Section~\ref{sec:histogramOfTimeToAlarm}).
\item An attrition adversary whose goal is to prevent consensus long
enough for random damage to corrupt the AU will be detected before
he succeeds (Section~\ref{sec:results:attrition}).
\end{itemize}

\begin{figure}
\centerline{\includegraphics{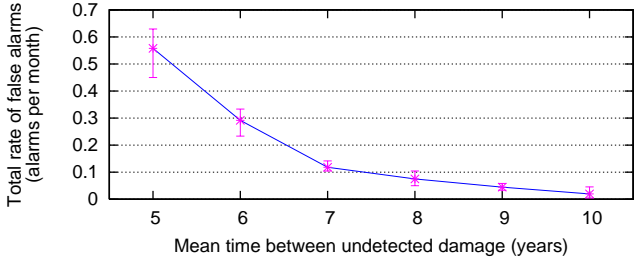}}
\caption{The total rate of false alarms versus the mean time between
random undetected damage at each peer, with no adversary.}
\label{fig:Mtbf}
\end{figure}

\subsection{Rate of false positives}
\label{sec:rateOfFalsePositives}

Without an adversary, but with peers subject to random damage they do
not themselves detect, Figure~\ref{fig:Mtbf} shows that false alarms
occur rarely.
We simulate 20 years with \emph{every} peer suffering random
undetected damage at mean intervals varying from 5 to 10 years.
Over 20 runs, we show the minimum, average and
maximum total rates of false alarms raised at \emph{any} peer in the
entire system.  With undetected damage at each peer every 5 years, in
the worst case the average rate of false alarms in the system is 44
days, that is, every 44 days some peer in the system sees an alarm.  The
average peer sees an alarm once in about 120 years.

The rates of random undetected damage we simulate are vastly higher than we
observe in practice.  Our peers typically lack reliability features such
as ECC memory.  Yet in over 200 machine-years of the test deployment,
we have observed only one peer in which such errors affected polls.  Our
simulations below assume this 1 in 200 probability of a random undetected
error per peer year.

\subsection{Stealth Adversary}
\label{sec:results:stealthAdversary}

We show that the probability of a stealth adversary causing
irrecoverable damage remains very low even for an initial subversion of
1/3, and then increases gradually.  Conservatively, we deem damage
irrecoverable if the initially subverted (malign) and the damaged
(loyal) peers form more than 50\% of the population.  For the following
simulations, the adversary has infinite CPUs and as many NICs as
necessary to gain the maximum possible foothold ratio during 20 years of
lurking.  We vary churn from 2\% to 10\% and subversion from 1\% to
40\%.  For every initial subversion and churn factor, we run all
compatible attack phases lasting up to 10 years for all foothold ratios
(40\% and up) achieved during the lurking runs (see
Section~\ref{sec:stealthSimulation}).  We run every combination of
parameters described above with 20 different random seeds.

\begin{figure*}
\centerline{\includegraphics{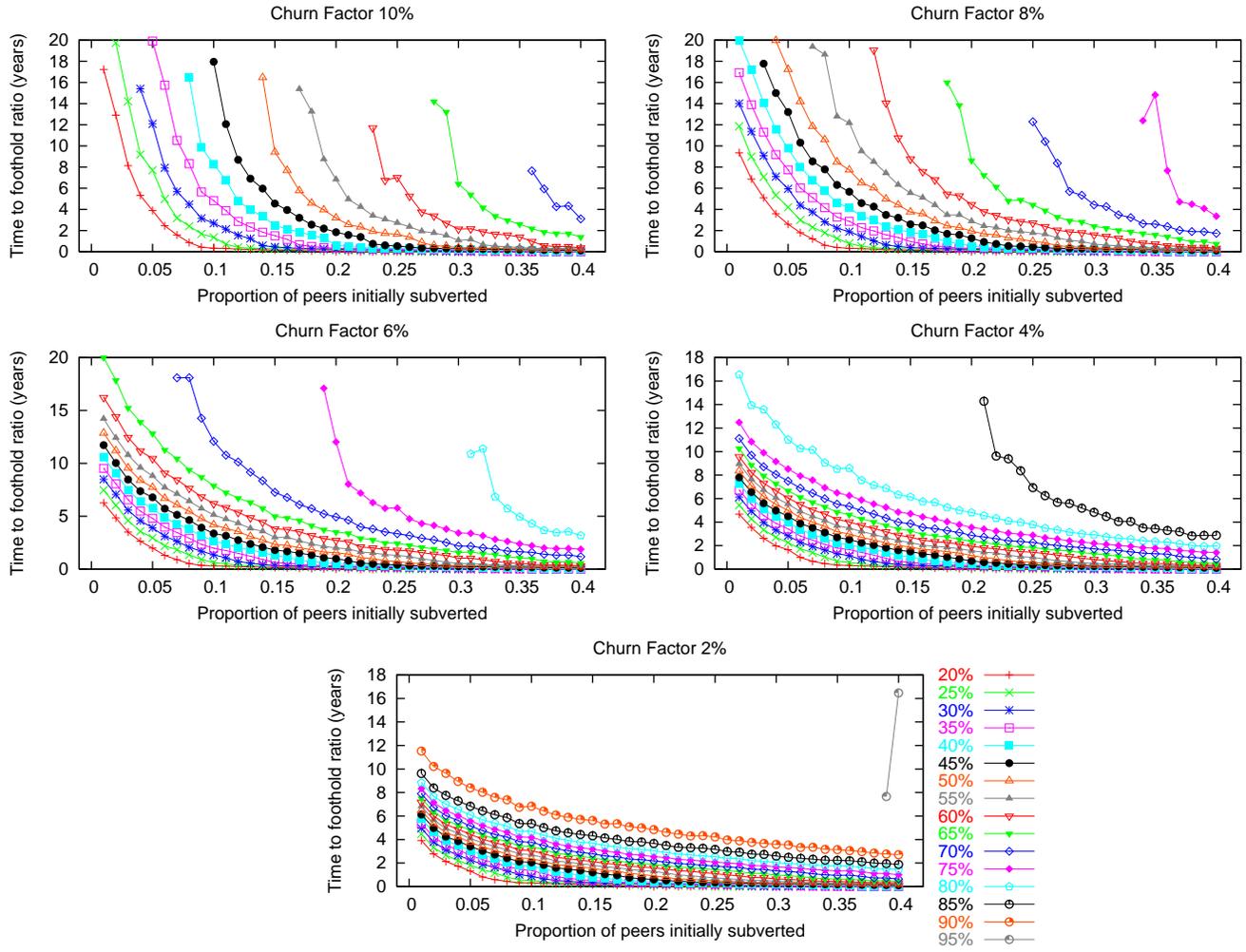}}
\caption{Minimum time for the stealth adversary to achieve various
foothold ratios, starting with various proportions of initially
subverted peers.  We show graphs for churn factors 2, 4, 6, 8, and
10\%.}
\label{fig:lurkPhase}
\end{figure*}

Figure~\ref{fig:lurkPhase} shows the minimum time taken by the lurking
phase to a foothold.  The $x$ axis shows the proportion of total peers
that are initially subverted.  The $y$ axis shows the minimum time it
takes the adversary to deteriorate the loyal peers' reference lists to
various foothold ratios.  Note that runs with low subversion levels do
not achieve foothold ratios of 20\% or more in 20 years.  Also, the
adversary can achieve greater footholds when loyal peers use a lower
churn factor.  This is because reference list churning resists the
attempts of the adversary to create a foothold in reference lists that
is much greater than the initial subversion he has obtained; less
churning means that the adversary can expand his foothold more
effectively.

\begin{figure*}
\centerline{\includegraphics{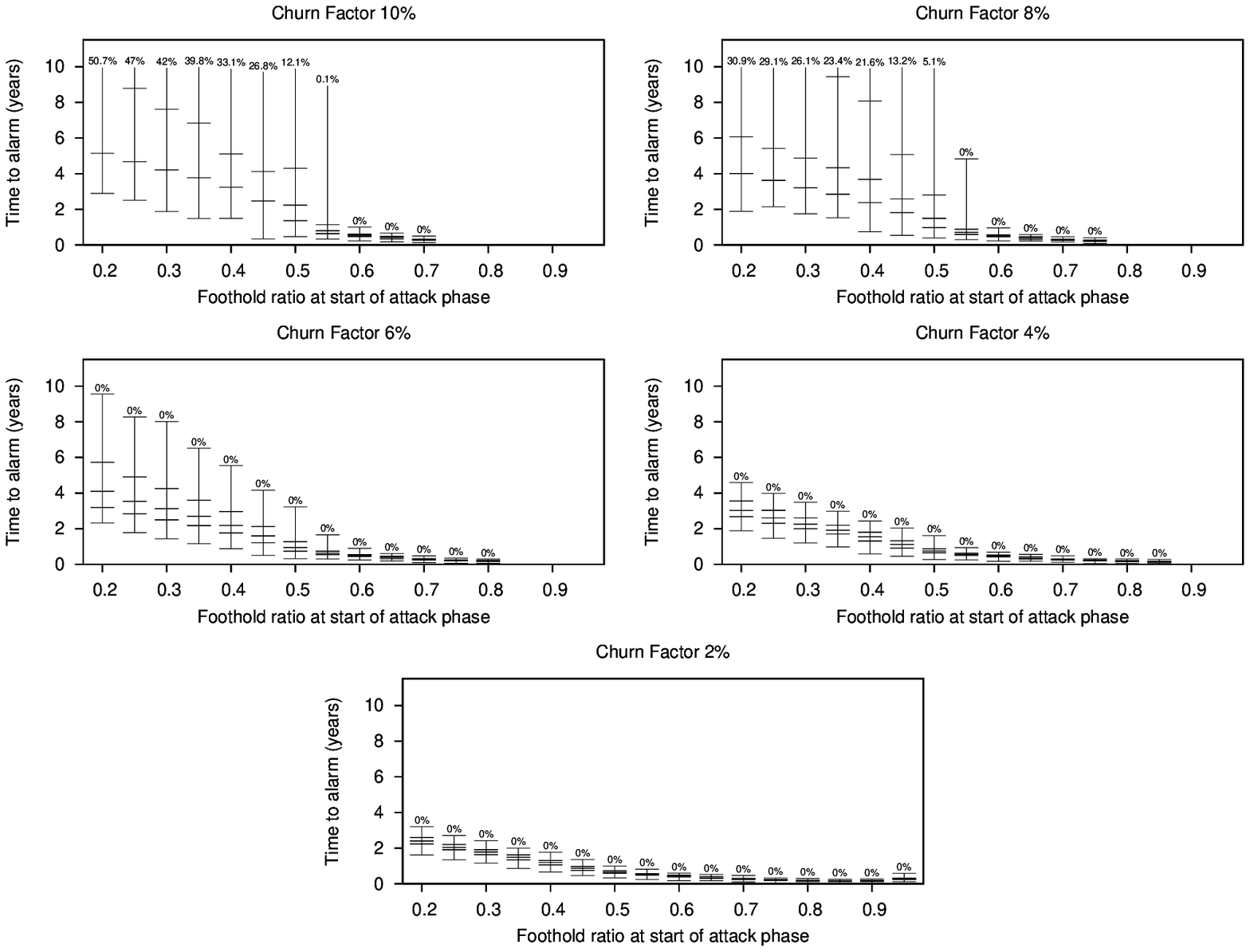}}
\caption{The time from the start of the attack phase (in the stealth
strategy) to the time of detection, for different starting reference
list foothold ratios.  Ticks split the value distributions into
quartiles.  Percentages above the distributions indicate runs that did
not generate an alarm.  We show graphs for churn factors 2, 4, 6, 8, and
10\%.}
\label{fig:attackPhaseAlarm}
\end{figure*}

Figure~\ref{fig:attackPhaseAlarm} shows how long the attack phase lasts
before it is detected.  For each foothold ratio at the beginning of the
attack phase, we show the quartile distribution of times until the first
alarm.  Some runs do not raise an inconclusive poll alarm; they damage
very few loyal peers.  At the top of each distribution is the percentage
of such runs.

\begin{figure*}
\centerline{\includegraphics{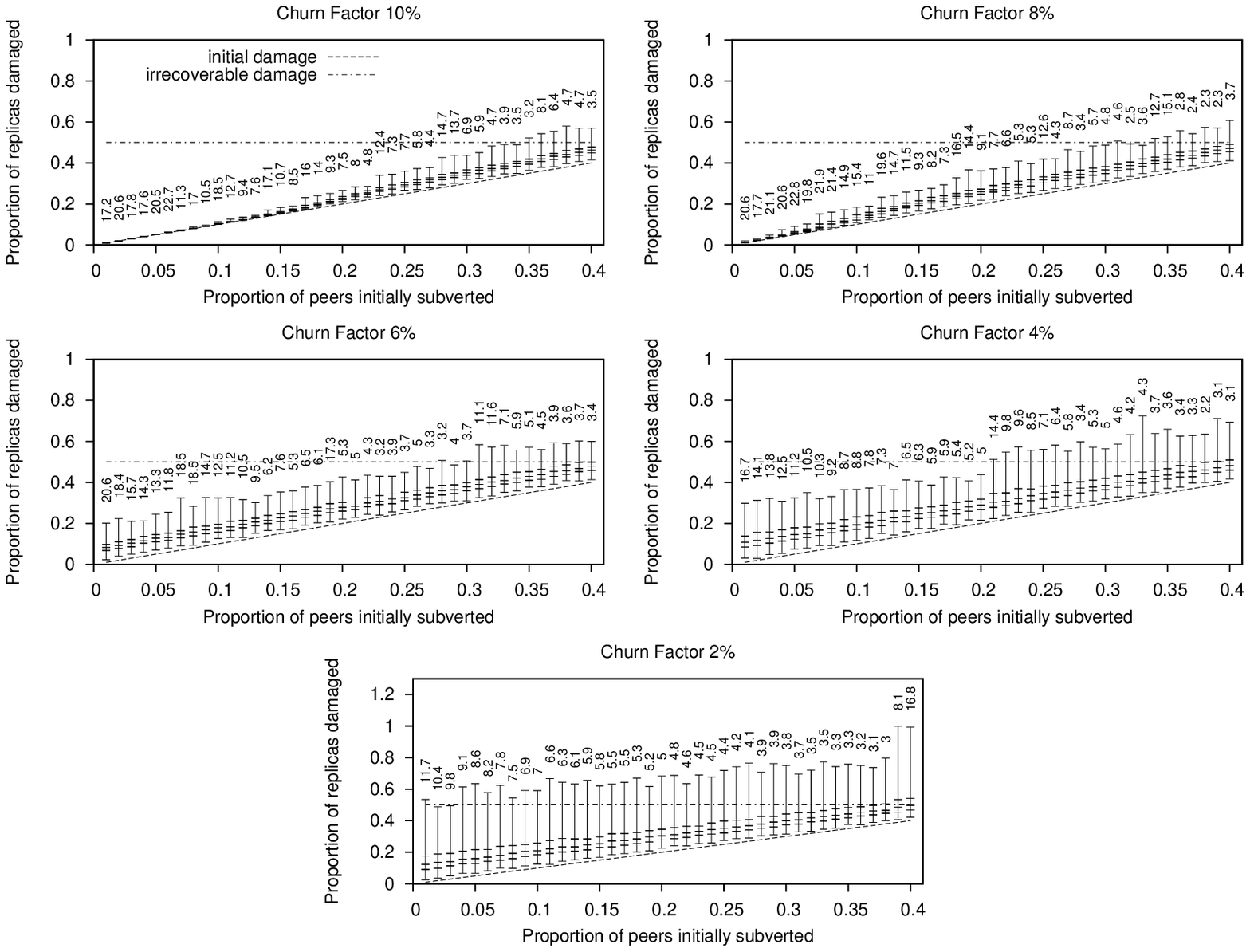}}
\caption{The percentage of bad replicas as a function of initial
subversion.  Ticks split the value distributions into quartiles.
Numbers above the distributions show the time in years needed by the
adversary to cause maximum damage.  The diagonal line shows the damage
due to peer subversion.  The horizontal line shows the threshold for
irrecoverable damage.
\label{fig:damagePercentile}}
\end{figure*}

\begin{figure*}
\centerline{\includegraphics{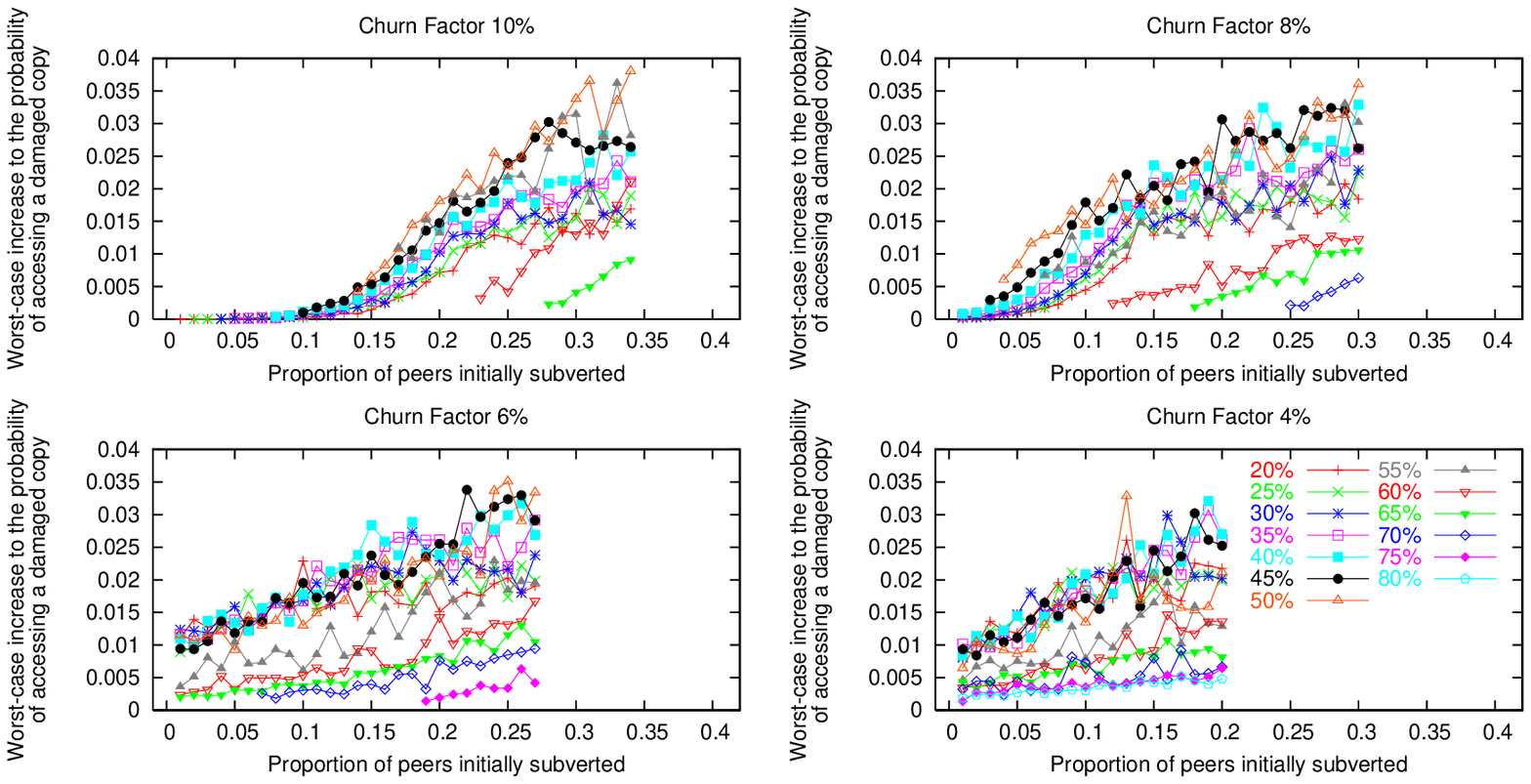}}
\caption{Worst-case increase in the probability of accessing a bad
replica due to the attack, as a function of initial subversion.  Each
curve corresponds to runs with a different foothold ratio at the start of the
attack phase.  We only show values for subversions at which the
adversary did not cause irrecoverable damage.  We show graphs for churn
factors of 4, 6, 8 and 10\%.
\label{fig:attackSummary}}
\end{figure*}

\begin{figure}
\centerline{\includegraphics{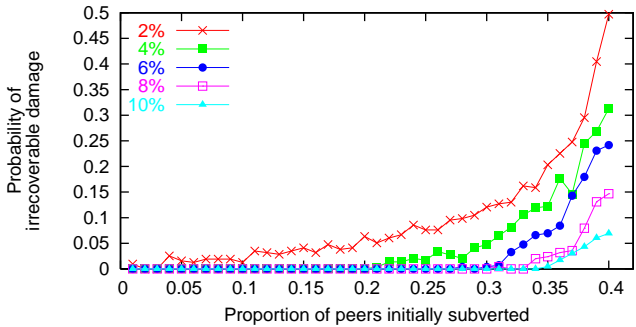}}
\caption{Probability that the adversary causes irrecoverable damage.
Each curve corresponds to a different churn factor.
\label{fig:adversarySuccesses}}
\end{figure}

Figure~\ref{fig:damagePercentile} illustrates that the adversary must
subvert a significant number of loyal peers to change the AU
irrecoverably for a churn factor of 10\%; the initial subversion need be
lower for lower churn factors.  The graph shows the distribution of the
maximum proportion of bad replicas (including those at subverted peers)
caused as a function of initial subversion.  The number above the
distribution shows the time in years needed to achieve the maximum
damage with that subversion.  For example, with a 10\% churn factor and
an initial subversion of 8\%, the adversary takes more than 16 years of
lurking to achieve a 40\% foothold ratio and another year of attacking
to damage an additional 0.8\% of the replicas.  As subversion increases,
the adversary is able to damage more loyal peers.  With a 10\% churn
factor and up to 34\% subversion, the adversary does not cause
irrecoverable damage; at 35\% subversion and above he succeeds in no
more than 12\% of runs (see the 10\% churn curve of
Figure~\ref{fig:adversarySuccesses}).

Figure~\ref{fig:attackSummary} summarizes the effect on readers of
attacks, isolating the benefit that the adversary obtains with the
strategy from what he is given through the initial subversion.  We omit
subversions for which irrecoverable damage is possible; in those cases, the
transient effect of the attack on readers is irrelevant compared to the permanent
loss of content that the adversary causes.

On the $x$ axis, we show the initial subversion.  On the $y$ axis, we
show the worst-case probability (due to the attack) that a reader of the
AU finds a damaged copy, i.e., the expected value of the maximum
fraction of damaged (not initially subverted) AUs during the lurk and
attack phases.  We graph a curve for each foothold at which the
adversary starts the attack phase, and we show a graph for each of churn
factors 4, 6, 8 and 10\% .  Interestingly, the adversary's best strategy
is not to lurk for as long as possible: readers are most likely to see a
bad AU when he lurks up to a foothold ratio of 50\% at 10\% churn;
lurking less results in weaker attack phases; lurking more means that
peers supply readers with good AUs for longer.  This behavior is
consistent also for the mean probability (not shown).  This ``best''
foothold ratio is lower for lower churn factors, e.g., a 40\% foothold is
best for the 4\% churn factor.  This is because the system's resistance
to the adversary's attempts to gain a foothold in reference lists is
weaker when churn is lower, i.e., when fewer friends are brought into
the reference list to temper the adversary's biased statistics.

Note also that for a churn of 10\%, despite an initial subversion of more
than 1/3 of the peers (34\%) by an adversary with unlimited
computational power, unlimited identities, complete knowledge of the
protocol parameters and an attack lasting more than a year, the
probability of a reader accessing a bad AU is only 2.7 percentage points
greater than it is immediately after the initial subversion.  The system
resists further damage effectively despite the subversion of 1/3 of its peers.

\begin{figure}
\centerline{\includegraphics{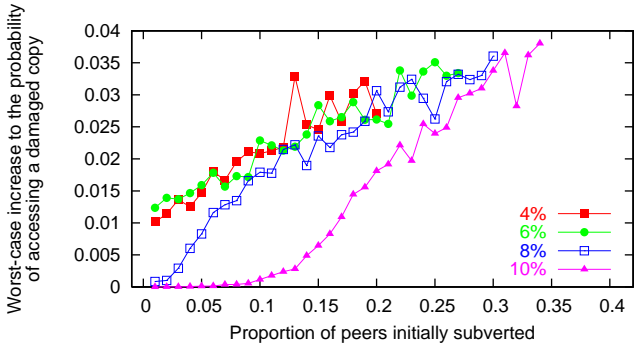}}
\caption{Worst-case increase in the probability of accessing a bad
replica due to the attack, as a function of initial peer subversion.
Each curve corresponds to runs with different churn factors.  We only
show values for subversions at which the adversary did not cause
irrecoverable damage, for each churn.
\label{fig:churn}}
\end{figure}

In Figure~\ref{fig:churn}, we explore how different churn factors affect
the worst-case probability of accessing a bad AU, over all foothold
ratios at the start of the attack phase.  Thus the curve for the 10\%
churn factor is the upper envelope of Figure~\ref{fig:attackSummary}.
We only show data for subversions at which irrecoverable damage
does not occur; this eliminates runs with 0 and 2\% churn factors, as
well as other points above a critical subversion (e.g., 23\% for 4\%
churn).  The graph shows that increasing the churn factor raises the
initial subversion the adversary needs before he can cause irrecoverable
damage, and reduces the probability of accessing a bad AU replica.

Finally, Figure~\ref{fig:adversarySuccesses} shows that churning the
reference list is an invaluable tool in thwarting the adversary.  On the
$y$ axis, we show the probability that the adversary causes
irrecoverable damage in our simulations, given different initial
subversions on the $x$ axis, one curve per churn factor.  Increasing the
churn factor increases the initial subversion needed to make
irrecoverable damage possible.  Beyond this critical subversion level,
the system suffers a gradual increase in the probability of
irrecoverable damage.  Note, however, that even with low churn factors
and up to 40\% initial subversion, the adversary still has a 37\% chance
of being caught before he causes irrecoverable damage.

\subsection{Nuisance Adversary}
\label{sec:histogramOfTimeToAlarm}

\begin{figure}
\centerline{\includegraphics{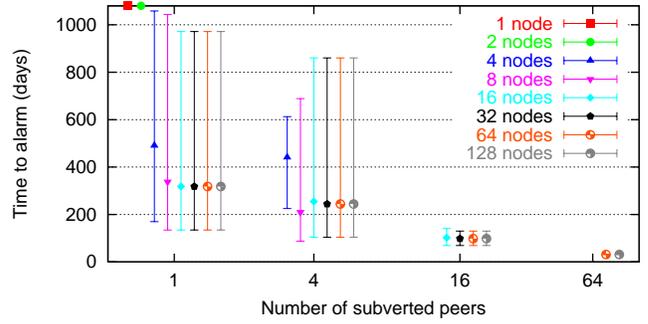}}
\caption{Time from start of nuisance attack to first alarm at any peer
against number of peers subverted,
for varying adversary resources.
Some attacks with few subverted peers do not cause alarms in our timescale.}
\label{fig:Nuisance}
\end{figure}

Figure~\ref{fig:Nuisance} shows the effect of a nuisance adversary
with 1 to 128 nodes-worth of computing effort available who
subverts 1 to 64 peers then attempts to raise an alarm at \emph{any} peer.
The simulation ends after 3 years
or at the first such alarm.
The error bars show minimum,
average and maximum times to the first alarm,
over 12 runs per data point,
with some attacks not generating an alarm in the first 3 years.

If the nuisance adversary subverts only a few peers, irrespective of his
computing resource, he takes about 6 months to raise an alarm, or (see
Figure~\ref{fig:Mtbf}) the equivalent of a random damage rate of once
every 6-7 years.  If he takes over a large number of peers, irrespective
of his computing resource the alarm happens quickly.  This seems to be
suitable behavior, in that large compromises should be detected.

\subsection{Attrition Adversary}
\label{sec:results:attrition}
 
\begin{figure}
\centerline{\includegraphics{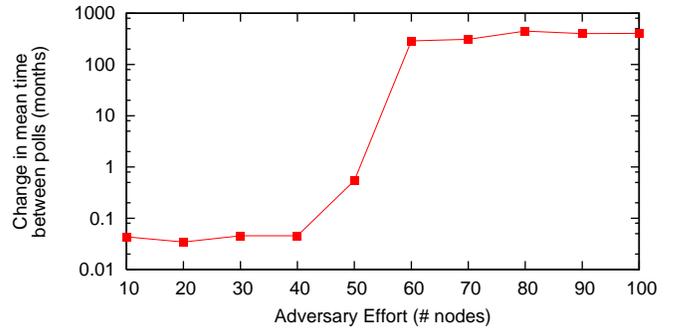}}
\caption{Effect of the attrition adversary in increasing mean time
between polls as a function of his effort.  The $y$ axis is in
logarithmic scale.}
\label{fig:attritionSummary}
\end{figure}

Figure~\ref{fig:attritionSummary} shows the increase in the average
inter-poll time at loyal peers as a function of the effort expended by
the attrition adversary.  Without sufficient resources, he has little
effect on the system.  If he can deploy sufficient resources, in our
case about 60 machines, he can prevent polls from reaching quorum and is
detected after 3 average inter-poll intervals (9 months), which is much
less than half any reasonable mean time between random undetected damage
at a peer.  Any damage accumulated can therefore be repaired after the
response to the alarm suppresses the attack.
 
If the attrition adversary with few resources focuses his attention on a
single peer he can cause it to raise an alarm after about 9 months.
This might be an effective way to deny service; further work is needed
to prevent it.

\section{Related Work}
\label{sec:related}
In common with the Byzantine-fault-tolerance (BFT) literature (e.g.,
\cite{Castro1999}, \cite{Chor1989},
\cite{Malkhi1998}, \cite{Reiter1994}), our voting protocol derives an
apparent prevailing opinion among a set of peers, some of whom are
malicious.  There are many differences; our population size is too large
for BFT's global communication, we degrade gradually rather than mask
the effects of failure or attack, and because we cannot assume an
absolute upper bound on the malicious peers' resources we have to
consider the possibility of being overwhelmed.  We use sampling to avoid
global knowledge or communication, rate-limiters to prevent our
adversary's unlimited resources from overwhelming the system quickly,
and integrated intrusion detection to preempt unrecoverable failure.

Our work has similarities with the anti-entropy protocol forming part of
Bimodal Multicast~\cite{Birman1999}, a reliable multicast protocol in
which peers send digests of their message histories to randomly chosen
other peers.  Peers receiving these messages can detect omissions and
request repairs from the peer that sent the digest.  The system's name
comes from its bimodal distribution of delivery probability, which is
similar to our distribution of poll results absent an attack.  As in our
case, it exploits the properties of random graphs.  As the authors
acknowledge, the lack of voting among the peers leaves the anti-entropy
protocol vulnerable to malign peers.

Our work also shares a goal with some of the first peer-to-peer systems
including Freenet~\cite{Clarke2000}, FreeHaven~\cite{Dingledine2000},
and the Eternity Service~\cite{Anderson1996}, namely to make it hard for
a powerful adversary to damage or destroy a document in the system.  The
other key goal of these systems is to provide anonymity for both
publishers and readers of content, which we do not share.  It would make
our system both illegal and unworkable, since we often preserve content that
must be paid for.

Several studies have proposed a persistent, peer-to-peer storage service
including Intermemory~\cite{Chen1999}, CFS~\cite{dabek01},
Oceanstore~\cite{Kubiatowicz2000}, 
PAST~\cite{rowstron01c}, and Tangler~\cite{mazieres01}.  Some
(e.g., Oceanstore) implement access control by encrypting the data and
thus do not solve our preservation problem, merely reducing it from
preserving and controlling access to the content, to preserving and
controlling access to the encryption key.  Others (e.g., PAST) implement
access control based on long-term secrets and smartcards or a key
management infrastructure.  Neither is appropriate for our application.
Some (e.g., Intermemory) use cryptographic sharing to proliferate $n$
partial replicas, from any $m<n$ of which the file can be reconstituted.
Others (e.g., PAST) replicate the entire file, as we do, but do not allow
control over where the replicas are located.  The goal of the LOCKSS
system is to allow librarians to take custody of the content to which
they subscribe.  This requires that each library keep its own copy of a
document it has purchased, not share the load of preserving a small
number of copies.

Moore et al.~\cite{Moore2001} report interesting measurements on packet-level
denial-of-service attacks, types, and frequencies.  For example, most
attacks are relatively short with 90\% lasting less than an hour.
Rarely do such attacks span multiple days.  The relevance of this data
to application-level denial-of-service attacks is questionable, but our
simulated attacks require attention spans from the attackers several
orders of magnitude longer.

\section{Future Work}
\label{sec:future}
We have two immediate goals: to deploy an implementation to our test
sites, and to improve the protocol's performance against the attrition
adversary.  Before deployment, we need to emulate the initial version's
handling of many practical details, especially the
``divide-and-conquer'' search for damage in an AU formed of many
documents.  Our simulated attrition adversary can currently prevent 1000
loyal peers from running polls with about 60 malign nodes
fully committed to the attack.  We have identified
several avenues for improving this inadequate performance, including
using additional peer state to identify attrition attacks and improving
the economic model to account for commitment of time as well as
effort~\cite{Rosenthal2003}.

Our current adversary model starts by making a proportion of the peers
malign; these peers remain malign for the duration.  Other peers may
have their AUs damaged by an attack, but they remain loyal for the
duration.  We need to enhance the model to account for peers becoming
malign as they fall victim to vulnerabilities, and becoming loyal as
their administrators patch and repair them.

We have analyzed potential adversaries from two points of view, their
goals and the opportunities offered them by their roles in protocol
implementations.  But we have yet to show that our simulated adversary
strategies are optimal.  We have shown that the system's cost structure
and other parameters can be set appropriately for a realistic
application, but we have yet to explore alternate cost structures and
parameter values, or how to set them optimally.  These investigations
will be aided when we are able to validate the simulator against a
deployed implementation.

\section{Why Lots Of Copies Keep Stuff Safe}
\label{sec:discussion}

Storage alone will not solve the problem of digital preservation.
Academic materials have many enemies beyond natural bit rot: ideologies,
governments, corporations, and inadequate budgets.  It is essential that
sound storage and administration practices are complemented with the
formulation of communities that act together to thwart attacks
that are too strong or too extrinsic for reliable storage to
withstand.

In a novel thrust towards this goal, we have built a new opinion poll
protocol for LOCKSS, applying the design principles in
Section~\ref{sec:designPrinciples}:

\emph{Cheap storage is unreliable.}  We replicate all persistent storage
across peers, audit replicas regularly and repair any damage
they find.  Peer state is soft and rebuilt by normal system operations
if it is lost.

\emph{No long-term secrets.}  Our peers need keep secrets only for the
duration of a single poll.  Without long-term secrets attackers may
spoof the identity of peers, but by requiring evidence of recent effort
we reduce the time during which stability of identity matters to a few
poll durations,
and we use short-term secrets to reduce spoofing during a poll.

\emph{Use inertia.}  We provide the system with an analog of inertia by
making \emph{all} of its operations inherently expensive and by limiting
the rate of possible change in the system.  Because even the operations
involved in failure are inherently time-consuming, it is very hard for
attackers to overwhelm the system quickly, which provides time for
humans to prevent the attack from resulting in catastrophic failure.

\emph{Avoid third-party reputation.} Third-party reputation is
vulnerable to slander and subversion of previously reliable peers,
especially in the absence of strong identities.  Further, we do not use
accumulated evidence of a peer's good behavior.  We instead require
evidence of substantial recent effort to allow a peer to influence the
outcome of a poll.

To the extent to which a peer does maintain a history of another peer's
behavior, that history is very simple,
derived from direct observation,
and acts only as a hint.
The system survives the loss or corruption of this memory at peers.

\emph{Reduce predictability.}
The reference lists are the only mechanism by which an attacker can
damage a loyal peer's AU.  Churning deprives the attacker of complete
control of them,  slowing an attack and increasing the risk of
detection.

\emph{Intrusion detection is intrinsic.}
Random damage to individual replica AUs stored by peers is incoherent,
resulting in polls that are either landslide agreement or
landslide disagreement.
An attacker attempting to change a preserved AU requires
coherent damage to replica AUs across peers,
which results in polls that are more closely contested.
Contested polls,
and thus coherent damage,
raise an inconclusive poll alarm
and lead to detection of the attack.

\emph{Assume a strong adversary.}  We allow for an adversary who can
call on unlimited computing resources and unlimited identities, who
can subvert or spoof a large proportion of the peers, and who has
information about the parameters of each poll that a real attacker
would be unlikely to possess.

\section{Conclusion}
\label{sec:conclusion}

We have shown that a combination of massive replication, rate
limitation, inherent intrusion detection and costly operations can
produce a peer-to-peer system with remarkable ability to resist attacks
by some extraordinarily powerful adversaries over decades.  Its lack of
dependence on long-term secrets and stable identities blocks many of the
paths by which systems are typically attacked.  We believe that this
protocol will allow us to scale the deployed LOCKSS system to production
levels in 2004.  Although we developed the new LOCKSS protocol for an
application with unusual characteristics, especially its goal of
preventing change, we nonetheless believe that the concepts and
principles underlying the protocol will be useful in the design of other
long-term large-scale applications operating in hostile environments.

The LOCKSS project is hosted by SourceForge, where the current
implementation can be obtained.  The Narses simulator will be available
from SourceForge shortly.  Both carry BSD-style Open Source licenses.

\section{Acknowledgments}
\label{sec:acknowledgments}
We have benefited from discussions with colleagues who
provided helpful feedback, test code and early paper manuscripts on
related work or tools.  In particular, we would like to express our
gratitude to Cynthia Dwork, Kevin Lai, Aris Gionis, Katerina Argyraki,
Lada Adamic, Sergio Marti, Ed Swierk, Hector Garcia-Molina, and the
LOCKSS engineering team and beta sites. Vicky Reich, the director of the
LOCKSS program, immeasurably helped us stay on track and remain
nourished before deadlines.  Finally, the anonymous SOSP reviewers and
John Heidemann, our paper shepherd, provided many pertinent and
constructive comments.

This work is supported by the National Science Foundation (Grant No.\
0205667), by the Andrew W. Mellon Foundation, by Sun Microsystems
Laboratories, by the Stanford Networking Research Center, by DARPA
(contract No.\ N66001-00-C-8015), by MURI (award No.\
F49620-00-1-0330), and by Sonera Corporation.  Any opinions, findings,
and conclusions or recommendations expressed here are those of the
authors and do not necessarily reflect the views of these funding
agencies.

\bibliographystyle{plain}
\bibliography{../common/bibliography}

\appendix
\section{Economic Considerations}
\label{sec:economicDesign}
We use a memory-bound function (MBF) scheme due to Dwork et
al.~\cite{Dwork2003}.  Here we briefly describe it and compute
appropriate costs to impose on poll initiation and voting.  Recall that
$S$ is the size of the AU in cache lines.

\subsection{Overview of a Memory Bound Function
Scheme}
\label{sec:overviewDwork}
The goal of the MBF is to cause the \emph{prover} of the necessary
effort to incur a number $C$ of cache misses and thus RAM accesses.  If
each of these takes $t$ seconds, the prover must have used $C \cdot t$
seconds on a real computer.  Memory bandwidths vary significantly less
among commonly available architectures than CPU speeds do, making MBFs
superior in provable effort to the CPU-bound functions previously
proposed~\cite{Dean2001,Dwork1992}.

The scheme we use has two adjustable parameters, the cost, $l$, of verifying
an effort proof and the ratio, $E$, between $l$ and the cost of
constructing the proof.  We measure all costs in cache misses, so a
proof costs $E \cdot l$ cache misses to construct and $l$ cache misses
to verify.

Dwork et al.\ describe an MBF scheme that uses an incompressible fixed
public data set $T$ larger than any cache it is likely to meet.  In our
case, a gigabyte would be practical.  An effort prover who must expend
effort $E \cdot l$ is given as challenge a nonce $n$ (so that the prover
cannot reuse older effort proofs) and the values of $E$ and $l$.  In
response, the prover must perform a series of pseudo-random walks in the
table $T$.  For each walk, the prover starts from a different position
$s$ of his choosing and computes a one-way value $A$ based on $n$, $s$
and the encountered elements of table $T$.  The walk is dependent on $n$
and $s$; it is constructed so that the number of encountered elements is
$l$, and fetching each encountered element causes an L1 cache miss. Each
walk, therefore, causes $l$ cache misses.

The prover stops when he computes a value $A$ that has 0 bits in its
least significant $\log_2E$ positions.  With the given MBF scheme, it
is expected the prover will try $E$ walks with different starting
positions $s$ before finding an appropriate starting position $s'$; this
costs the prover $C = E \cdot l$ cache misses.

The $s'$ that yielded the appropriate $A$ is the effort proof.  The
verifier need only perform the random walk on $T$ starting with the $n$
he chose and the $s'$ sent by the prover; this costs the verifier $V =
l$ cache misses.  If the resulting value $A$ from this walk has the
proper 0-bits in its last $\log_2E$ positions, the verifier accepts the
effort proof as valid.

We first describe how we use the MBF scheme in the construction and
verification of votes.  We then describe how we impose costs on vote
construction and verification and how we choose appropriate parameters
for each of these.

\subsection{Vote Construction and Verification}
\label{sec:VoteConstructionAndVerification}

\subsubsection{Construction}
\label{sec:appendixVoteConstruction}

As described in
Sections~\ref{sec:protocol:specification:voteVerification} and
\ref{sec:protocol:specification:voteConstruction}, vote construction is
divided into rounds.  Each round consists of two parts: the construction
of an MBF proof of effort and the hashing of a portion of the document.
The portion of the document hashed in each round has twice as many
content blocks as the previous round, $2^{i-1}$ blocks in round $i$, or
$r = \lceil \log_2(B + 1) \rceil$ rounds for a content of $B$ blocks (we
explain why in Section~\ref{sec:votingEffortSizes}).

We design vote construction in a way that ensures the order of
computation for each stage; specifically, we wish to ensure that voters
cannot precompute or parallelize different stages of the construction of
a vote (although individual stages of MBF proof construction may be
parallelizable~\cite{Dwork2003}).

A vote contains the MBF proofs and content hashes computed during the
two stages of every vote construction round.  We denote the list of MBF
effort proofs for the $r$ rounds of vote construction as $[s_1, \ldots,
s_r]$ and the list of the corresponding content hashes as $[H_1, \ldots,
H_r]$.

At round $i$, the MBF proof computation stage takes as input the nonce
$n_i$ (we explain how we build this nonce below) and the proof
parameters $E_i$ and $l_i$, returning the proof $s_i$ and the output
value $A_i$ (see Section~\ref{sec:overviewDwork}).  This is followed by
the content hash stage of round $i$, which computes the hash $h(s_i \|
A_i \| \mathit{content-block}_i)$, where $\|$ denotes bit string
concatenation, $h$ is our cryptographic hash function, and
$\mathit{content-block}_i$ denotes the portion of the content that we
hash during round $i$ (to be determined in
Section~\ref{sec:votingEffortSizes}).  The output of the hashing stage
is the hash value $H_i$.

Because the proof and output of the MBF proof computation are included
in the input to the cryptographic hash function $h$, the vote
constructor cannot precompute $H_i$ before having determined the MBF
proof $s_i$ and the corresponding output value $A_i$.  We include both
$s_i$ and $A_i$ in the hash (as opposed to only the proof $s_i$) to
ensure that it is hard for the vote constructor to precompute $H_i$
values for all possible $s_i$s.  Instead, $s_i$ and $A_i$ together come
from a large enough range of possible values that precomputing all
likely hashes $H_i$ by brute-force is intractable.

The nonce $n_i$ input into the MBF proof effort computation of the
$i$-th round must be such that precomputing an effort proof for round
$i$ before the MBF proof effort computation or content hashing stage of
round $i-1$ is intractable.  For the first round of vote construction,
$n_1$ must be such that the voter cannot start computing its vote until
it receives the challenge of the poll initiator.  As a result, $n_1 =
h(\mathit{challenge} \| \mathit{pollID} \| \mathit{voterAddress})$, and
for $i > 1$, $n_i = h(s_{i-1} \| A_{i-1} \| H_{i-1} \|
\mathit{challenge} \| \mathit{pollID} \| \mathit{voterAddress})$.

\subsubsection{Verification}

Vote verification is also divided into rounds, with MBF and hashing
stages in each round.  The initiator, guided by the list of MBF effort
proofs and hashes contained in the \msg{Vote} it receives, verifies the
computations of the voter using its local copy of the document.

At round $i$, the MBF proof verification stage takes as input the nonce
$n_i$ (see Section~\ref{sec:appendixVoteConstruction}), the proof
parameters $E_i$ and $l_i$, and the proof $s_i$ (included in the
message) and constructs the output value $A_i$ (see
Section~\ref{sec:overviewDwork}).   If this value ends in $\log_2E_i$
0-bits, the verifier accepts the effort proof.  Otherwise, the verifier
deems the vote invalid.

If the verifier has yet to deem the vote disagreeing, it proceeds with
the content hashing stage of the vote verification.  Specifically, it
computes the hash of its appropriate local content blocks: $H_i' = h(s_i
\| A_i \| \mathit{content-block}_i)$.  If the resulting hash $H_i'$ is
different from the value $H_i$ contained in the vote, the verifier deems
the vote disagreeing and only verifiers MBF efforts in the remaining
rounds.

Note that, as with vote construction, the verifier sets $n_1$ using the
challenge, poll identifier and the voter's identity, and all subsequent
$n_i$'s using, again, the challenge, poll identifier and voter's
identity, as well as the effort proof $s_{i-1}$ from the vote, the value
$A_{i-1}$ computed during the previous round's MBF stage, and the
\emph{voter's} hash $H_{i-1}$ from the vote.

\subsection{Choice Of Voting Effort Sizes}
\label{sec:votingEffortSizes}

Let $C_\mathit{MBF}(i)$ and $V_\mathit{MBF}(i)$ be the costs of the
$i$-th round MBF construction and verification, respectively.  If there
are $b$ cache lines in a content block, then the $i$-th round hashing
costs $H(i) = 2^{i-1} b$ and is the same in construction and in
verification.  Then, round $i$ vote construction costs $C_v(i) =
C_\mathit{MBF}(i) + H(i)$ and vote verification costs $V_v(i) =
V_\mathit{MBF}(i) + H(i)$.  Finally, overall voting operations cost $C_v
= \sum_{i=1}^{r} C_v(i)$ for construction, and $V_v = \sum_{i=1}^{r}
V_v(i)$ for verification.  The cost to verify a disagreeing vote may be
less, since hashing stops after the first disagreeing block.

We require (Section~\ref{sec:protocol:analysis:effortSizing}) that the
cost of constructing a vote be greater or equal than the cost of
verifying that vote, even if the vote is corrupt or malformed, in that
in round $i$ either its MBF verification fails or its content hash
disagrees.  We examine the four cases in which our requirements must
hold: valid agreeing vote, valid disagreeing vote, a vote that contains
garbage from the $i$-th MBF proof onwards, and a vote that contains
garbage from the $i$-th content hash onwards.  We set $E_i = E$,
constant for all rounds, and set $l_i$ to the size of the content block
in each round, that is $l_i = 2^{i-1}b$.  Different choices for $l_i$
and variable $E$ for each voting round are possible.

\subsubsection{Valid Agreeing Vote}

The verifier performs both the MBF and content hashing components of
every verification round.  As a result, we must establish that
$C_v \geq V_v$:
\begin{align}
&& C_v                           & \geq V_v \nonumber \\
\Leftrightarrow && \sum_{i=1}^{r} C_v(i)            & \geq \sum_{i=1}^{r} V_v(i) \nonumber\\
\Leftrightarrow && \sum_{i=1}^{r} C_\mathit{MBF}(i) & \geq \sum_{i=1}^{r} V_\mathit{MBF}(i) \nonumber\\
\Leftrightarrow && \sum_{i=1}^{r} El_i             & \geq \sum_{i=1}^{r} l_i \nonumber\\
\Leftrightarrow && \label{eqn:validAgreeing}E      & \geq 1\
\end{align}

\subsubsection{Valid Disagreeing Vote}

If we satisfy the inequality~\ref{eqn:validAgreeing} for agreeing votes,
we fulfill the requirements for disagreeing votes, because if the
content hash disagrees in round $i$, no further hashing is performed.
Thus the cost of verifying in this case is no greater than $V_v$.

\subsubsection{Invalid $i$-th MBF Proof}

If the vote's MBF verification fails in round $i$ but not before, its
constructor must have performed at least the effort of the preceding
rounds: $C_{v,\mathit{invMBF}}(i) = \sum_{k = 1}^{i - 1}{C_v(k)}$.  The
verifier's effort ceases after the MBF verification of round $i$, and
costs no more than $V_{v,\mathit{invMBF}}(i) = V_\mathit{MBF}(i) +
\sum_{k = 1}^{i - 1}{V_v(k)}$.  The economic requirements of our
solution impose that $C_{v,\mathit{invMBF}}(i) \geq
V_{v,\mathit{invMBF}}(i)$:
\begin{align}
&& C_{v,\mathit{invMBF}}(i)                               & \geq V_{v,\mathit{invMBF}}(i) \nonumber \\
\Leftrightarrow && \sum_{k = 1}^{i - 1}{C_v(k)}           & \geq V_\mathit{MBF}(i) + \sum_{k = 1}^{i - 1}{V_v(k)} \nonumber \\
\Leftrightarrow && \sum_{k = 1}^{i - 1}(C_v(k) - V_v(k))  & \geq V_\mathit{MBF}(i) \nonumber \\
\Leftrightarrow && \sum_{k = 1}^{i - 1}\
                   (C_\mathit{MBF}(k) - V_\mathit{MBF}(k))& \geq V_\mathit{MBF}(i) \nonumber \\
\Leftrightarrow && \sum_{k = 1}^{i - 1}(El_k - l_k)       & \geq l_i \nonumber \\
\Leftrightarrow && b(E - 1)\sum_{k = 1}^{i - 1}2^{i-1}    & \geq b2^{i-1} \nonumber \\
\Leftrightarrow && (E - 1)(2^{i-1} - 1)                   & \geq 2^{i-1} \nonumber \\
\Leftrightarrow && E(2^{i-1} - 1)                         & \geq 1 \nonumber \\
\Leftrightarrow && E                                      & \geq \frac{1}{2^{i-1} - 1}\
\label{eqn:cheatMBF}
\end{align}
This requires that $2^{i-1} > 1$, which holds for all $i > 1$.

For the first voting round ($i = 1$), the inequality is unsatisfiable.
Even if the malicious voter sends garbage, the verifier
must perform some effort,
the first round's MBF verification,
to detect it.
Sending garbage costs the verifier $V_\mathit{MBF}(1) = b$.
But to be invited to send garbage,
the malicious peer had to be in the initiator's reference list,
which cost at least $C_v = (E + 1) S$.
Sending garbage squanders this effort,
$(E+1)$ times the hash cost of the entire content,
to impose a blocks' worth of cache misses.

\subsubsection{Invalid $i$-th Content Hash}

If the $i$-th round content hash fails to agree,
the vote is certainly disagreeing, but may still be valid.
The verifier deems it invalid only when the next round MBF verification fails.
In this case, the vote construction must have cost the voter at least
$C_{v,\mathit{invHash}}(i) = C_\mathit{MBF}(i) + \sum_{k = 1}^{i -
  1}{C_v(k)}$ and 
the vote verification costs the verifier
$V_{v,\mathit{invHash}}(i) = V_\mathit{MBF}(i+1) + \sum_{k = 1}^{i}{V_v(k)}$.
We require
$C_{v,\mathit{invHash}}(i) \geq V_{v,\mathit{invHash}}(i)$:
\begin{align}
&& C_{v,\mathit{invHash}}(i)                         & \geq V_{v,\mathit{invHash}}(i) \nonumber \\
\Leftrightarrow && C_\mathit{MBF}(i) + \
                \sum_{k = 1}^{i - 1}{C_v(k)}         & \geq V_\mathit{MBF}(i+1) + \
                                                           \sum_{k = 1}^{i}{V_v(k)} \nonumber \\
\Leftrightarrow && Eb2^{i-1} + \
                \sum_{k = 1}^{i - 1}(C_v(k) - V_v(k)) & \geq b2^{i}+ V_v(i) \nonumber \\
\Leftrightarrow && Eb2^{i-1} + \
                \sum_{k = 1}^{i - 1}(El_k - l_k)    & \geq b2^{i} + 2b2^{i-1} \nonumber \\
\Leftrightarrow && Eb2^{i-1} + \
                (E-1)b\sum_{k = 1}^{i - 1}2^{k-1}   & \geq 4b2^{i-1} \nonumber \\
\Leftrightarrow && E2^{i-1} + (E-1)(2^{i-1} - 1)    & \geq 4 \cdot 2^{i-1} \nonumber \\
\Leftrightarrow && E(2\cdot 2^{i-1} - 1)            & \geq 5 \cdot 2^{i-1} - 1 \nonumber \\
\Leftrightarrow && E                                & \geq \frac{5 \cdot 2^{i-1} - \
                                                                 1}{2 \cdot 2^{i-1} - 1}\
\label{eqn:cheatHash}
\end{align}
provided that $2^i > 1$, which is true for all $i > 0$.

\subsubsection{Choice for $E$}

Any choice of $E$ that satisfies Inequalities~\ref{eqn:validAgreeing},
\ref{eqn:cheatMBF} and \ref{eqn:cheatHash} for $i > 1$ is appropriate
given our economic requirements.  Note also that different choices for
$l_i$ and even variable $E$ for each voting round are also possible.
For our choice of $l_i$, $E = 4$ is the minimum value for $E$.
Higher values might be desirable for smaller content sizes to
make rate limitation more effective.
If $S$ is the size of the content in cache lines and $E = 4$,
the cost of constructing a vote is $C_v = 5S$,
the cost of verifying an agreeing vote $V_v = 2S$
and the cost of verifying a disagreeing vote $V_\mathit{vd} = S$.

\subsection{Designation Of Poll Initiation
  Effort Sizes}
\label{sec:pollingEffortSizes}
We want the cost $C_p$ of poll initiation per invitee to be at
least the induced cost on the invitee:
\begin{equation}
\label{eqn:pollingCost}C_p \geq V_p + C_v
\end{equation}
where $V_p$ is the cost of verifying the poll initiation effort.

Based on the analysis of the voting costs above, this means
\begin{align}
&& E_p l_p &\geq l_p + 5S \nonumber \\
\Leftrightarrow && (E_p - 1) l_p &\geq 5S
\end{align}
One choice for the MBF parameters is $E_p = 4$ and 
$l_p = (5/3)S$.  The poll initiator must expend 
$C_p = (20/3)S$ cache
misses per invitee, and each invitee must spend 
$V_p = (5/3)S$ cache
misses to verify the invitation.

\end{document}